\documentclass{jfm}
\usepackage{graphicx}
\usepackage{epstopdf, epsfig}
\usepackage{amsmath}
\usepackage{here}
\usepackage{comment}
\usepackage{lmodern}
\usepackage[dvipsnames]{xcolor}
\usepackage{placeins}
\usepackage{float}
\usepackage{hyperref}
\usepackage{caption}
\usepackage[mathscr]{euscript}

\title{Effect of transient shell formation on Shock-induced atomization of an evaporating nanofluid droplet}

\author{
    Gautham Vadlamudi\aff{1},
    Bal Krishan\aff{1}
    Akarsh Choudhary\aff{1}
    \and Saptarshi Basu\aff{1}
        \corresp{\email{sbasu@iisc.ac.in}}
    }

\affiliation
{\aff{1} Department of Mechanical Engineering, Indian Institute of Science, Bangalore, India.
}

\begin{document}

\maketitle

\begin{abstract}
Understanding the breakup of evaporating nanofluid droplets under extreme aerodynamic loads is essential for applications in drug delivery, material deposition, defence, and propulsion. This study investigates the shock-induced atomisation dynamics of an acoustically levitated TM-10 nanofluid droplet subjected to a coaxially propagating blast wave and subsequent compressible vortex ring, generated using a compact wire-explosion shock source. The blast wave imposes a sharp velocity discontinuity, followed by a decaying flow field and a vortex-dominated interaction that drives droplet disintegration.

Laser-induced heating promotes evaporation, increasing nanoparticle concentration, viscosity, and agglomeration within the droplet. Progressive evaporation leads to interfacial nanoparticle accumulation, initiating a sol–gel transition when the local volume fraction exceeds the gelation threshold, and ultimately forming a solid outer shell as the maximum packing limit is approached.

Shock interactions are systematically examined across three evaporation stages: (i) steady liquid phase, (ii) gel-shell phase, and (iii) solid-shell phase. Each regime exhibits distinct atomisation responses due to the evolving interfacial morphology. In the gel-shell regime, deformation is resisted by the viscous shell, producing a bag-on-sheet mode, followed by puncture, jetting, and eventual shell delamination. In the solid-shell regime, interactions intensify, resulting in brittle fragmentation and catastrophic fracture.

The findings reveal how evaporation-driven interfacial transitions fundamentally alter breakup mechanisms under transient shock loading. By linking nanoparticle transport, shell formation, and flow–droplet interaction across multiple timescales, this work establishes new physical insights into the atomisation of complex, multicomponent, multiphase, and transiently evolving droplets under extreme aerodynamic conditions.
\end{abstract}


\section{Introduction}\label{sec:intro}

Understanding the principles of droplet atomisation is integral to advancing a wide spectrum of engineering technologies. This ubiquitous phenomenon of atomisation and externally imposed flow underpins critical processes such as combustion dynamics, evaporative cooling, agricultural spraying, aerosol lubrication, liquid metal atomisation for powder production, and spray drying/coating \citep{eggers2008physics, bayvel2019liquid, kapilan2023comprehensive, sharma2023shock, cervantes2014study,krishan2021efficacy,vadlamudi2022penetration,vadlamudi_insights_2023,vadlamudi2021insights}. A liquid droplet subjected to the stream of gas show different modes of breakup, each mode leading to different breakup mechanism and secondary droplet sizes \citep{xu2023transitions, hopfes2021secondary}. Accurately predicting and controlling the physics of aero-breakup is critical for optimising the performance and efficiency of the aforementioned processes \citep{guildenbecher2009secondary}.

Extensive research has been carried on the dynamics of droplet breakup, with studies covering both incompressible and compressible flow regimes, utilising the air-jet flow \citep{chou1997temporal,chou1998temporal,dai2001temporal, zhao2013temporal}, shock tubes \citep{theofanous2004aerobreakup}, and blast wave \citep{sharma2021shock, chandra2023shock}. Several numerical studies have also been carried out to understand the dynamics of droplet deformation and breakup in the cross flow \citep{aalburg2003deformation, meng2015numerical, sembian2016plane, jain2015secondary}.
It is well established that the breakup behaviour of droplets is primarily determined by the physical properties of the fluid and the characteristics of the surrounding flow. This behaviour is typically characterized using two dimensionless numbers: the Weber number (\( \textrm{We} \)) and the Ohnesorge number (\( \textrm{Oh} \)), which are defined as
\begin{align}
    \textrm{We} = \frac{\rho U^2 D_0}{\sigma}, \quad \textrm{and} \quad
    \textrm{Oh} = \frac{\mu}{\sqrt{\rho \sigma D_0}},
\end{align}
where \( \rho \) is the density of the droplet, \( U \) is the relative velocity between the droplet and the surrounding fluid (typically air), \( D_0 \) is the characteristic length scale (usually the droplet diameter), \( \gamma \) is the surface tension, and \( \mu \) is the dynamic viscosity of the droplet. When viscous effects are small ($Oh<0.1$), classical experiments have identified several breakup regimes that occur due to the variation in $We$. In particular, five canonical modes (with We ranges) have been presented by several researchers, where $We < 8$ the droplet undergoes vibrational mode of breakup.
At \(We \sim 10 - 25\), the droplet undergoes \emph{bag breakup}, where it deforms into a thin, inflated bag that eventually bursts. As the Weber number increases to $We \sim 25-65$, a \emph{bag-stamen (plume) breakup} occurs, characterised by the formation of a bag with a trailing tail (jet like) of liquid. In the range of \(We \sim 65-85\), a \emph{multibag breakup} is observed, where successive, nested bags form over the flattened droplet surface, which break into tiny secondary droplets. At even higher Weber numbers, from 85 to 120, the droplet experiences \emph{sheet-thinning/stripping} breakup, in which a broad equatorial liquid sheet is continuously stripped from the droplet. Finally, at \(We\) values greater than 350, the droplet undergoes \emph{catastrophic breakup} due to high shear effects, where it suddenly shatters into numerous tiny fragments \citep{cao2007new, jackiw2021aerodynamic, sharma2021shock}. Each mode is driven by fluid instabilities: low-$We$ bag-type deformations arise from Rayleigh–Taylor instabilities at the droplet front, while high-$We$ sheet breakup is dominated by Kelvin–Helmholtz (KH) surface waves that form holes and ligaments \citep{zhao2020breakup}. Numerous reviews have compiled these regimes and transition criteria \citep{pilch1987use, guildenbecher2009secondary, theofanous2011aerobreakup} to guide spray design and fundamental studies. However, \cite{sharma2021shock} has reclassified the breakup regimes, specifically for shock-induced aerobreakup of droplets, showing Rayleigh-Taylor piercing (RTP) mode induced bag/multibag breakup for low \(We<100\) and KH-induced shear stripping dominantly occurring for \(We>1000\). Furthermore, the in-between \(We \approx 10^2–10^3\) shows mixed mode of atomisation, where both Rayleigh-Taylor (RT) and Kelvin-Helmholtz (KH) instabilities are present the the breakup mode depends on the growth rate of these different types of instabilities. The KH-instabilities result in the formation of a liquid sheet that subsequently perforates through hole nucleation and ligament ejection \citep{sharma_shock_2021}. 

Recent high-speed experiments and simulations have deconstructed the shock–droplet interaction into two distinct stages \citep{sharma_shock-induced_2023}. The first stage involves complex shock wave dynamics at the droplet interface, including transmission, reflection, and diffraction, which instantaneously alters the pressure and velocity fields. The second stage is the subsequent airflow-induced breakup, where the shear flow stretches the droplet, leading to the development of instabilities. These findings on impulsive loading are consistent with the classical view of droplet breakup, which also shows bag breakup at low \(We\) and sheet stripping at high \(We\). Recent studies on non-Newtonian droplets, such as that of \cite{chandra2023shock}, who subjected viscoelastic (polymeric) droplets to shock-induced flow, have identified similar trends, categorising the breakup into three modes: vibrational, shear-entrainment, and catastrophic. Their work also reported that elasticity plays a minimal role during the initial stages of droplet deformation.

This study focuses on the breakup dynamics of evaporating nanofluid (TM10) droplet using a blast wave. TM10 is an aqueous solution of dispersed nano particles of silica (SiO\textsubscript{2}), of size $\sim$ 22 nano-meters . Nanofluids are engineered colloids with suspended nanoparticles, typically ranging from 10 to 100 nm, that exhibit superior thermophysical properties compared to their base fluids. These enhanced properties include a notable increase in thermal conductivity, improved surface wetting characteristics, and a non-Newtonian viscosity often described by sol–gel behaviour \citep{li2022effect,phuoc2009synthesis,sefiane2008contact, wang1999thermal}. Upon evaporation TM10, it leaves accumulated particles behind, forming a semi-rigid shell. \cite{sanyal2016controlling,chen2010effects} showed that the evaporation rate measured by the $D^2$ vs $t$ plot, is governed by the nano-particle concentration, type of nano-particle and the surfactant used. Experiments show that droplets
with suspended nano-particles (NPS) form a dense surface shell as they evaporate. A recent study by \cite{krishan2024evaporation} also demonstrated that the presence of a non-volatile component in these multicomponent droplets leads to the formation of a shell due to the accumulation of non-volatile particles at the droplet surface. The morphology of this shell is controlled by the size and nature of the particles. The shell can be thought of as a semi-solid crust enveloping the liquid core. \cite{pandey2017combustion} showed that this shell also acts as a porous, gelatinous skeletal medium, that hinders the surface evaporation in nanofuels (Dodecane + Alumina NPs) resulting in reduction of droplet evaporation rate after the formation of shell. Continued evaporation within the shell drives the outward flow of the liquid core, thereby increasing the internal capillary pressure. Sessile droplet study by \cite{miles2023effect} showed that when the compressive stress acting on the shell surpasses a critical threshold, it undergoes mechanical failure, leading to buckling, fracturing, or bursting. \cite{tsapis2005onset} demonstrated this behaviour in colloidal silica drops: after an initial phase of isotropic shrinkage, the drop suddenly buckles like an elastic shell, which coincides with a transition of the surface layer from fluid-like to solid-like.

This shell formation profoundly alters the droplet’s response to external stresses such as aerodynamic loading. The shell originates from evaporation-driven gelation: as solvent evaporates, the local particle volume fraction $\phi$ increases, leading to a sol--gel transition \citep{zang2019evaporation}. Rheological measurements of colloidal silica dispersions show a transition from Newtonian fluid behaviour at $\phi < 0.3$ to a viscoelastic paste for $\phi \sim 0.5$, and eventually to a brittle elastic solid beyond $\phi \approx 0.5$ \citep{di2012rheological}. As $\phi$ approaches the glass-transition limit ($\phi \sim 0.58$), the viscosity diverges due to hydrodynamic crowding and structural arrest \citep{russel2013divergence, cheng2002nature}. Consequently, the initially deformable droplet develops a rigid crust that undergoes buckling, delamination, or fracture, once internal stresses exceed the shell strength \citep{style2011crust, tsapis2005onset}. These failure modes are driven by the coupling of internal solvent fluxes and external aerodynamic stresses.

Other factors also modulate this sol--gel transition, such as solvent type, ambient temperature, and particle morphology \citep{chen2007rheological,gaganpreet2015viscosity}. Beyond gelation, shell solidification is typically triggered when the local volume fraction approaches the maximum packing fraction $\phi_m$, but $\phi_m$ itself decreases under particle agglomeration, since aggregates act as larger effective particles with reduced packing efficiency \citep{mohtaschemi2014rheology,studart2007colloidal}. This framework connects directly to shell mechanics. \cite{style2011crust} emphasised how surface particle accumulation, driven by evaporation, generates a mechanically solid crust that thickens over time and eventually buckles or fractures under internal capillary stress. In practical terms, a thin shell that is initially deformable can become almost rigid as evaporation progresses. Post-gelation, once the structural arrest sets in, the shell resists smooth deformation, and imposed stresses, whether from internal capillarity or external airflow, leading to abrupt failures such as buckling, cracking, or delamination. In other words, a nanofluid droplet under aerodynamic loading ruptures in a brittle fashion, fundamentally different from the smooth deformation and breakup of a pure Newtonian droplet.  

The combined effects of gelation, shell formation, and aerodynamic loading yield a far more complex breakup response in nanofluid droplets. Rather than smooth deformation and progressive sheet rupture, nanofluid droplets may experience brittle shell cracking, asymmetric rupture, or large-fragment ejection. Despite its importance, this coupled problem remains poorly understood, especially with respect to aerobreakup, where most of the literature has mainly focused on Newtonian/complex fluids having uniform spatial properties \citep{sharma_shock-induced_2023}.

The present study seeks to bridge a critical gap by investigating the aerodynamic fragmentation of acoustically levitated, evaporating nanofluid droplets subjected to blast waves. It aims to unravel the coupling between evaporation-driven shell formation and mechanics with aero-breakup processes, thereby extending and redefining the current framework of droplet fragmentation to particle-laden systems. The experiments reported here are first-of-their-kind, providing a holistic examination of nanofluid droplet evaporation within the inherently unsteady and dynamic context of shock-induced aero-breakup and atomisation. This work addresses the transient nature of aero-breakup with its multiple competing response timescales, and systematically explores how shell mechanics at different stages of evaporation govern droplet disintegration and secondary atomisation.

By embedding shell mechanics into the framework of droplet fragmentation, this study also offers a new perspective directly relevant to a wide range of particle-laden or multi component and/or multiphase systems. These include nanofluid spray cooling in microelectronics and power electronics, slurry fuel atomisation in advanced combustion, spray pyrolysis for nanoparticle production, thermal barrier and nanoparticle coating processes, suspension and colloidal sprays in additive manufacturing and materials processing, inhalable therapeutic aerosols, agricultural sprays with nanoparticle additives, or capsule dynamics in food/cosmetic/bioengineering or pharmaceutical applications \citep{chang1993experimental,unverfehrt2015deformation,grandmaison2021modelling} and multiphase blast or explosion environments where particulate-laden droplets control dispersal, ignition, or deposition behaviour. The findings, therefore, provide a foundation for interpreting fragmentation in complex droplets that bridge fluid mechanics, materials science, and applied engineering.

This study primarily focuses on the phenomenological investigation, focusing on the of aerodynamic breakup of a nano-fluid droplet with gelatinous/solid shell-like structures on its surface, due to evaporation. As a first-of-its-kind investigation into the shell dynamics of shock-induced aero-breakup, the present study primarily aims to establish a foundational understanding and provide a phenomenological analysis of how different shell mechanics influence atomisation behaviour.

In this work, we first discuss the laser-induced evaporation characteristics of nanofluid (TM10) droplet. Mechanism of sol-gel transition and shell formation is also discussed (see \autoref{sec:sample_Char}). The \autoref{sec:Flow_field} details the flow field characteristics of the blast wave. A global observation is presented in the \autoref{subsec:global Obs} and the time scales of droplet breakup are compared with the flow time scales in \autoref{sec:Int_timescales}. Effect of shock delay on the atomisation dynamics is disussed in the \autoref{sec:atomz}, and finally the key outcomes of this study are presented in the \autoref{sec:conclusion}.

\section{Experimental Methodology}

In the present study, the atomization mechanism of an evaporating nanofluid droplet is investigated to understand the influence of evaporation on atomization behaviour. A droplet is acoustically levitated and heated using a laser beam to induce controlled evaporation. At a specific stage during the droplet’s evaporation lifetime, a blast wave (generated using a compact shock tube) is triggered to interact with the droplet, resulting in its atomization. The overall experimental setup schematic is shown in \autoref{fig:EXP_setup}.

\subsection{Experimental setup}\label{sec:Exp_setup}

\subsubsection{Shock generation apparatus}
A compact in-house shock generation apparatus has been used for studying the droplet atomisation. The shock generator operates on the principle of high-voltage wire explosion and has been extensively used for shock/blast generation by different researchers, particularly for droplet atomization studies \citep{sharma_shock_2021,chandra_shock-induced_2023,sharma_shock-induced_2023,vadlamudi_insights_2024}. When a high-voltage electrical pulse of the order of a few kilovolts is imposed across a thin metallic wire, it results in rapid joule heating of the wire, which leads to a blast wave formation due to rapid vaporisation of the wire \citep{oshima_blast_1962}. In the current system, a 2kJ power pulse generator (Zeonics Systech, India Z/46/12) that houses a 5$\mu$F capacitor is used to supply the high-voltage pulse across the two electrodes with a thin metallic copper wire (35 SWG, Length, $L_{w}$=81mm) in electrical contact with the electrodes. The charging voltage of the capacitor is varied between 4kV to 10kV, resulting in blast waves with different shock strengths. The Mach number of the blast wave is measured at the reference location where the droplet is acoustically levitated. In the current experiments, the reference Mach number ($M_{s,r} = ({u_{s}}/{c})_{ref}$) ranges from 1.05 to 1.35. Here, $u_{s}$ is the speed of the blast front as measured at its instant of interaction with the droplet, and $c$ is the local speed of sound under ambient conditions (Temperature of 298K and pressure of 1atm). The plot depicting the dependency of $M_{s,r}$ on the capacitor charging voltage is shown in \autoref{fig:5_FlowChar}b. 

A shock tube (cross-section $4cm \times 2cm$) is mounted centrally over the electrode base plate of the electrode chamber, over which the thin copper wire is placed in electrical contact with the electrodes. The whole shock-generation setup is secured on a rigid structure, in horizontal orientation.

\begin{figure}
   \centering
    \includegraphics[width=.96\textwidth]{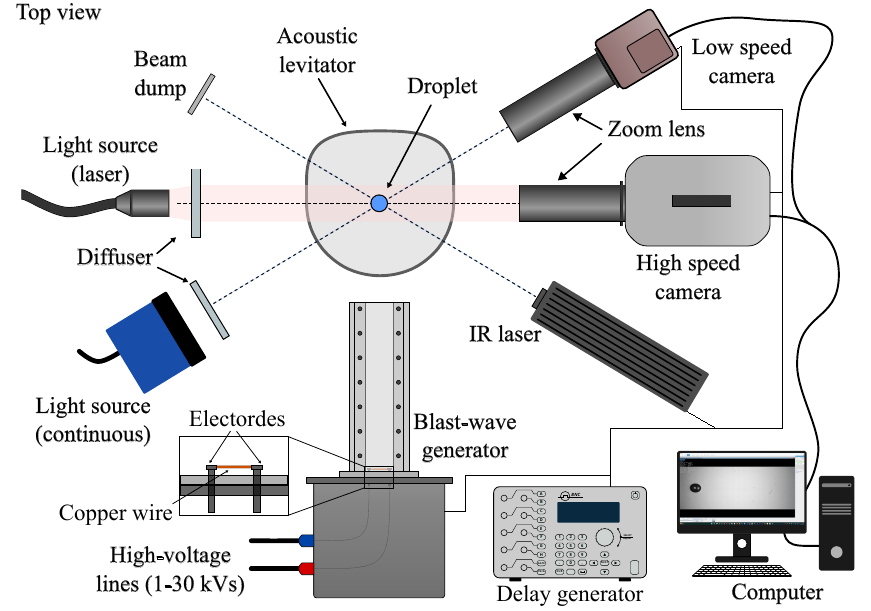}
   \caption{The schematic presents a top-down view of the experimental setup. The top view of the setup shows a droplet levitated using an acoustic levitator and heated using an IR laser. The blast wave setup is shown (below), which is used to generate different strengths of propagating blast wave by supplying different voltage to its electrodes. A pulsed laser and a continuous light source are used to illuminate the droplet to capture of breakup dynamics with a high-speed camera and slow evaporation behaviour with a low-speed camera.}
   \label{fig:EXP_setup}
\end{figure}

\subsubsection{Acoustic Levitation and droplet heating}  
A droplet of initial diameter ($D_0$) is levitated using the BIGLEV, a 3D-printed, multi-emitter, single-axis acoustic levitator comprising 72 transducers (16mm, 40kHz), following the work of \citep{marzo2017tinylev,hsu2020performance}. This levitator is mounted at a distance of 6cm from the shock tube opening, with the central node of the levitator aligned with the shock tube axis. The acoustic pressure of the levitator is controlled by controlling the voltage supplied to the levitator circuit (9 - 14V). A $CO_2$ infrared IR laser (Synrad 48, wavelength $\lambda = 10.6\mu m$) is used for externally heating the droplet due to the irradiation from the laser beam ($3.5mm$ diameter) having a maximum irradiation intensity of $I_{max}=1.04 MW/m^2$. The laser beam is aligned with the central node of the levitator where the droplet is levitated (see \cite{krishan2024evaporation}). The laser irradiation intensity of $I = 0.20 \times I_{max}$ has been used in current experiments, which has been chosen such that the heating rate is high enough for significant shell formation, as well as sufficient time is available for different stages of shell formation, but any internal nucleation is avoided.

\subsubsection{Nanofluid preparation} \label{sec:NF_preparation}
The Sigma-Aldrich LUDOX TM-40 colloidal silica i.e., aqueous solution of $40\%$ by weight 22nm silica ($SiO_2$) nanoparticles (NP) is diluted using Deionized (DI) water to obtain TM-10 (aqueous solution of $10\%$ by weight silica NPs). The weight percentage of the silica nanoparticles in the TM-10 nanofluid used in current experiments ($w_o \sim 0.1$) corresponds to a particle volume fraction of $\phi_o \sim 0.048$. 

\subsection{Experimental procedure}   

The following experimental procedure was followed to study shock-induced atomisation of a nanofluid droplet at a specific stage of evaporation, corresponding to a heating duration of $t_L=t_d$ (shock delay), by subjecting it to a specific shock strength (capacitor charging voltage). During each experimental run, a droplet is initially positioned at the central acoustic node of the levitator using an ultra-fine needle syringe (31G), and the $CO_2$ laser is set to a predetermined heating rate. Simultaneously, the 5$\mu$F capacitor is charged to a voltage level slightly higher than the required $kV$ to achieve a specific $M_{s,r}$. Once charged, the charging circuit is cut off, and the voltage starts to discharge continuously. A digital delay generator (BNC 575) is used to trigger all the different components in the experimental setup. As soon as the delay generator is triggered, a TTL signal is provided to the $CO_2$ laser (with pulse width of $t_d$ seconds) simultaneously, and another TTL signal is provided to the shock-generation system after $t_d$ seconds, which generates the blast wave by exploding the wire. The time gap \(t_d\) between the onset of evaporation and the wire explosion is hereafter referred to as the \emph{shock delay}.

\subsubsection{Imaging and data acquisition}
The interaction process was investigated using a combination of high-speed and low-speed shadowgraph imaging. For the imaging of droplet atomization, high-speed shadowgraph imaging has been performed at 40000 fps and 75000 fps using a High-speed star SA5 Photron Camera with navitar lens and Cavitar Cavilux smart UHS laser for backlighting. Simultaneous to the high-speed shadowgraph, low-speed imaging has been performed using an IDT color camera with a $1\times$ navitar lens and a continuous light source for backlighting to record droplet regression during evaporation. The spatial resolutions for high-speed and low-speed Shadowgraph imaging were set to 51.25 px/mm and 17.84 px/mm, respectively. The low-speed shadowgraph camera is triggered by the digital delay generator at $t_L=0$, simultaneously with the trigger of the heating laser, to record the slower droplet evaporation and regression data. The camera for high-speed recording is triggered along with the shock-generation system to record the high-speed phenomena of droplet atomization. 

For the flow characterisation of the blast wave system, a high-speed Schlieren setup is used to record the blast wave propagation and compressible vortex dynamics at 40000 fps and 75000 fps using a high-speed Star SA5 Photron Camera with Tokina lens. The Schlieren setup utilised a pair of parabolic concave mirrors with a focal length of 1500 mm with a light-source and a knife-edge placed at the focal points of the mirrors. A high-speed, non-coherent pulse diode laser (Cavitar Cavilux smart UHS, 400 W), emitting at a wavelength of 640 nm, served as the light source in the experiments (depicted in \autoref{fig:EXP_setup}). The spatial resolution for high-speed Schlieren visualisation had been set to 3.6764 px/mm.

\subsubsection{Data Processing} \label{sec:Data_Processing}
    
Schlieren imaging provided a means to visualise the evolution of the blast wave and CVR. By spatially tracking the position of the blast wave and CVR, with respect to time in Schlieren images, a time-dependent variation of the blast wave radius ($R_s$) was obtained. The data was then used to estimate the velocity of the blast front ($u_{s}$) and, subsequently, the blast wave Mach numbers ($M_{s}$). The width, height, and line-of-sight-area of the droplet has been evaluated using image processing and droplet detection from the shadowgraphy imaging. Otsu's thresolding algorithm has been used to binarise the image to detect the droplet boundary, to obtain the instantaneous droplet diameter ($D$).    

\section{Results and Discussions}\label{sec:results}

\subsection{Nanofluid Evaporation Characteristics}\label{sec:sample_Char}

\begin{figure}
    \centering
    \includegraphics[width=1\textwidth]{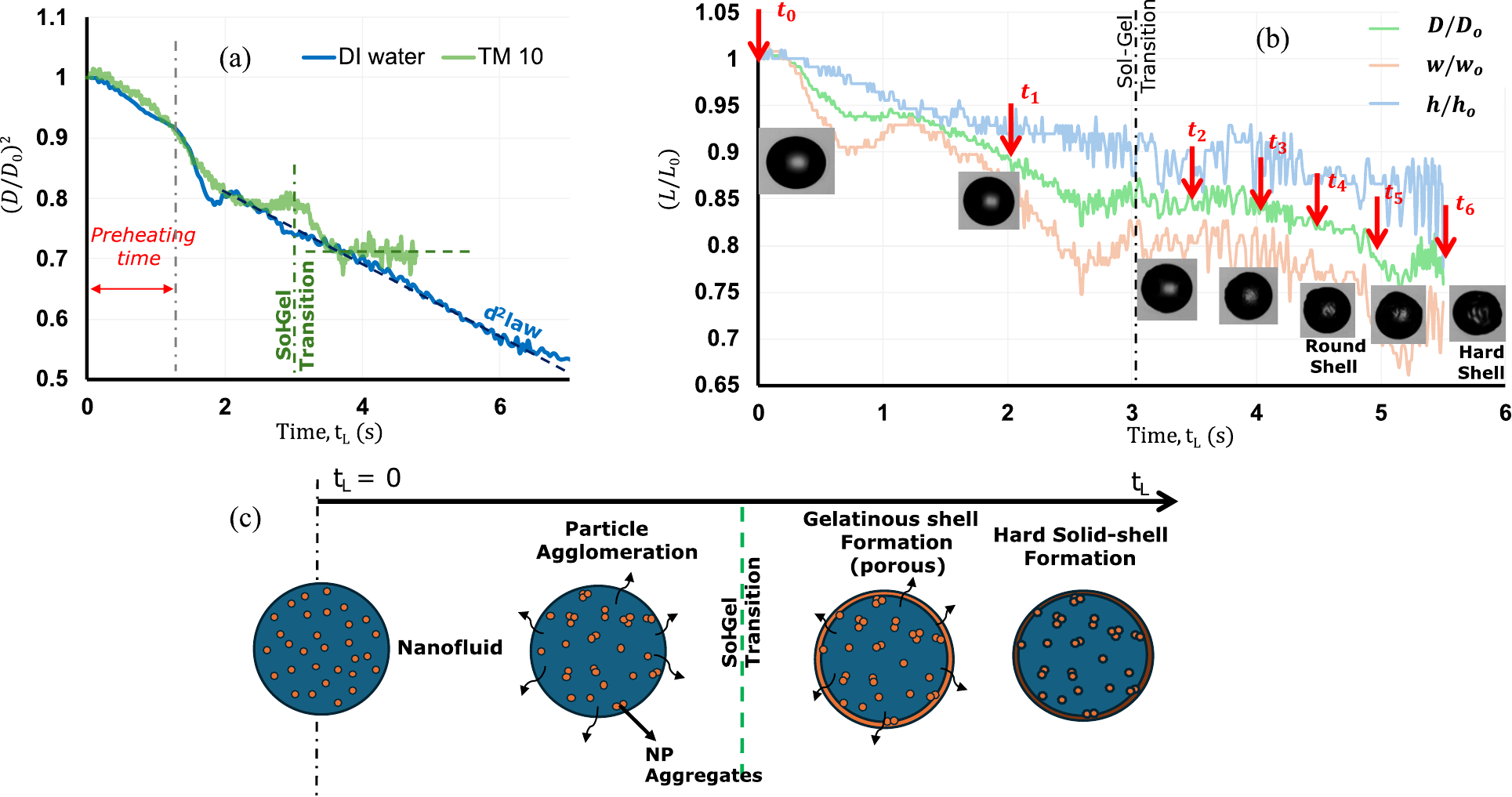}
    \caption{(a) Normalised droplet regression plot w.r.t. droplet lifetime ($(D/D_o)^2$ vs $t_L$) for DI water (blue) and TM-10 nanofluid (green), (b) Temporal variation of the normalised diameter, width and height of the TM-10 droplet, showing different stages of sol-gel transition and shell formation. The red vertical arrows indicate the different time delays from $t_L=0$ during droplet evaporation considered for the interaction with the shock flow, corresponding to the various stages of shell formation (refer to \autoref{tab:shock_Delays}). The snapshots of the droplet corresponding to the different stages during evaporation are shown. (c) Schematic depicting the interior of the nanofluid droplet, showing particle aggregation, sol-gel transition at the surface and shell formation, during the nanofluid evaporation.}
    \label{fig:1_stagesOfEvap}
\end{figure}

\autoref{fig:1_stagesOfEvap}a shows the temporal variation of the square of the normalised droplet diameter regression ($({D/D_o})^2$) during evaporation, plotted against the droplet lifetime ($t_L$). In Fig \autoref{fig:1_stagesOfEvap} the solid-blue line and solid-green lines represent the droplet regression for pure DI Water and nanofluid (TM-10) droplet. As shown by \cite{Pandey_Howboiling2018}, the droplet initially undergoes a preheating stage (see \autoref{fig:1_stagesOfEvap}a) during which the laser irradiation is utilised for the sensible heating of the droplet. Subsequently, the droplet regression occurs with a nearly constant surface temperature maintained at wet-bulb temperature ($T_{wb}$), where the square of the droplet diameter decreases linearly with time. This has been termed as the $D^2$ law in the literature, which occurs in the 'steady evaporation regime' during droplet evaporation. However, interestingly, nanofluid (TM-10) exhibited deviation from this behaviour after some time period as shown in \autoref{fig:1_stagesOfEvap}a(green line). The droplet size regression for TM-10 (green line) is observed to slow down and flatten around 4s, which corresponds to the porous skeletal shell formation at the surface of the droplet. \autoref{fig:1_stagesOfEvap}c
shows the schematic of the interior of the droplet depicting the shell formation mechanism.

As the nanofluid droplet evaporates, the nanoparticle (NP) concentration increases with time, and this also results in the time-dependent agglomeration of NPs forming larger aggregates, as shown in \autoref{fig:1_stagesOfEvap}c. As shown in the schematic, during the droplet regression, the NP aggregates get collected by the receding droplet surface, resulting in an increase in NP aggregate concentration at the droplet surface. This leads to the sol-gel transition near the droplet surface, when the particle volume fraction reaches a critical value at the surface. This results in the formation of a porous gelatinous shell at the droplet surface, represented by the orange layer in \autoref{fig:1_stagesOfEvap}c. After the gelatinous shell formation, the evaporation rate is reduced as the liquid has to evaporate through the porous shell \citep{PANDEY2019167}, resulting in a slower droplet regression rate (termed as "unsteady evaporation regime"). During the unsteady evaporation, as the droplet regression proceeds further, the agglomeration increases at the shell, resulting in the eventual solidification of the shell (hard-solid shell formation).

\autoref{fig:1_stagesOfEvap}b shows the temporal variation of the width, height and diameter of the nanofluid droplet during evaporation. The sol-gel transition shown with a dotted line in \autoref{fig:1_stagesOfEvap}b has been identified as the time at which the droplet diameter regression rate exhibits a significant reduction. In \autoref{fig:1_stagesOfEvap}b, the droplet width, height and diameter are observed to become nearly constant during the final stage ($t_L \sim 5s$), which corresponds to the hard-solid shell formation, where the shell becomes rigid after solidification. The various time delays ($t_d$) during droplet evaporation chosen for the investigation of the shock-induced atomization (in current experiments) are marked by red arrows in \autoref{fig:1_stagesOfEvap}b. $t_0$ corresponds to the unheated droplet case (ambient temperature), and $t_1$ to $t_6$ correspond to the evaporating droplet (with heating). $t_1$ and $t_2$ correspond to the steady evaporation regime and the sol-gel transition point, respectively. $t_3$ and $t_4$ correspond to the gelatinous shell regime (referred to as the ‘Round shell regime’), while $t_5$ and $t_6$ correspond to the solid shell regime (referred to as the 'Hard-shell regime'). For the generalisability of the different shock delays considered \((t_L \sim t_d = t_0,t_1,...,t_6)\), they are normalised by the gelation timescale \((t_g)\), defined as the time after which a gelatinous shell begins to form at the nanofluid droplet surface \citep{zang2019evaporation}. The gelation process will be discussed in detail in subsequent sections, and the normalised values corresponding to the different shock delays are summarised in \autoref{tab:shock_Delays}. As shown in \autoref{tab:shock_Delays}, values of \(t_d/t_g < 1\) correspond to the steady evaporation regime, whereas \(t_d/t_g > 1\) indicates the unsteady evaporation regime. Furthermore, the range \(1 < t_d/t_g \leq 1.5\) is associated with the gelatinous-shell regime, while \(t_d/t_g > 1.5\) corresponds to the solid-shell regime.

\begin{table}
    \centering
    \begin{tabular}{cc}
        \textbf{Shock delays}  &  \textbf{Normalised timescale} \\
        \textbf{($t_d$)} & \textbf{($t_d/t_g$)}\\
          &  \\
        $t_0$ &  0\\
        $t_1$ &   0.67\\
        $t_2$ &   1.17\\
        $t_3$ &   1.33\\
        $t_4$ &   1.5\\
        $t_5$ &   1.67\\
        $t_6$ &   1.83
    \end{tabular}
    \caption{Table of shock delays \((t_L \sim t_d = t_0, t_1, \dots , t_6)\) normalised by the sol--gel transition timescale \((t_g)\), expressed as \(t_d/t_g\). The table lists the corresponding normalised values for different shock delays considered in the present experiments, as illustrated in \autoref{fig:1_stagesOfEvap}.}
    \label{tab:shock_Delays}
\end{table}

 \subsection{Characterization of the flow-field induced by the blast wave} \label{sec:Flow_field}

\begin{figure}
    \centering
    \includegraphics[width=1\textwidth]{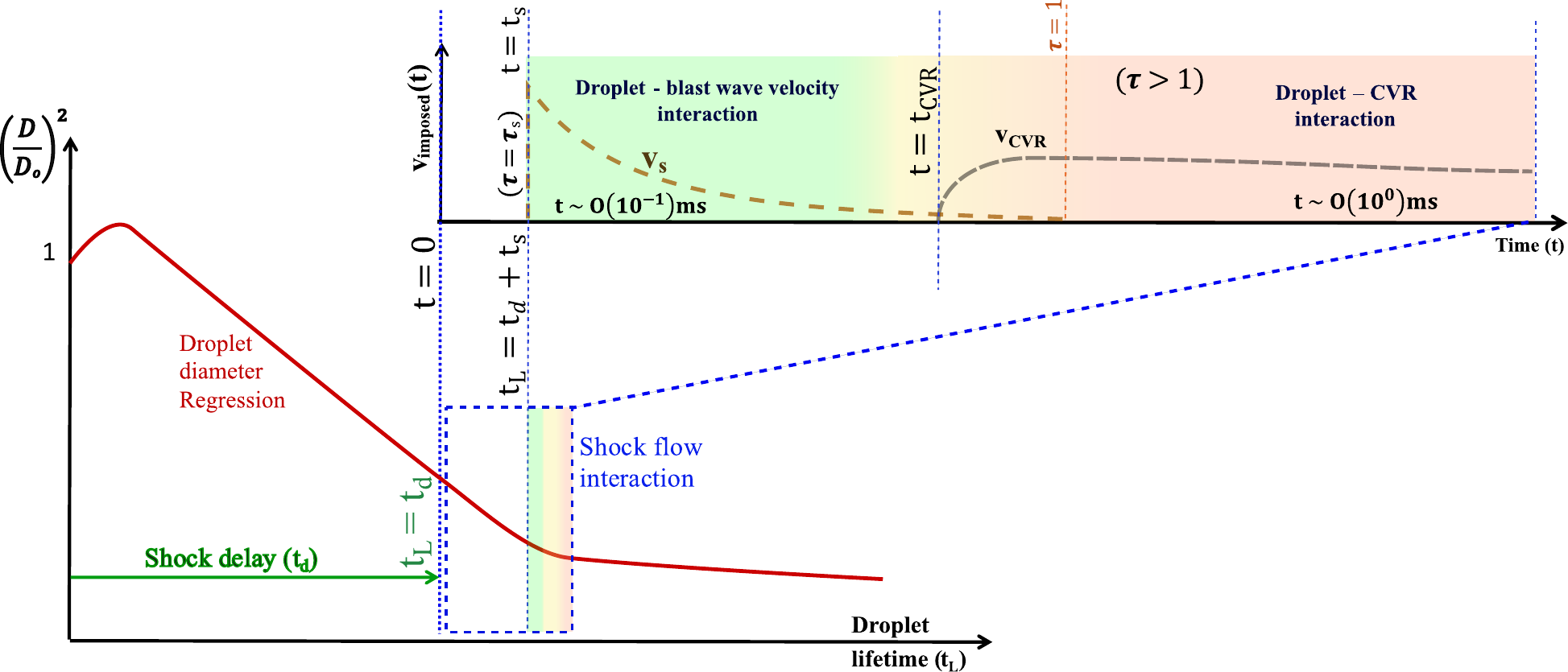}
    \caption{Stages of interaction: Schematic of the droplet regression (left) and Schematic of different stages of flow interaction with the droplet (top-right).}
    \label{fig:4_Int_stages}
\end{figure}

 The plot at the bottom of \autoref{fig:4_Int_stages} shows the schematic of the diameter regression of the evaporating nanofluid droplet. Here, $t_L$ represents the evaporation timescale of the droplet, where droplet heating is initiated at $t_L=0$. The solid red line in $(D/D_0)^2$ vs $t_L$ plot represent the temporal variation of normalised droplet diameter. In the current experiments, a blast wave is generated by exploding the wire and is triggered after a delay of $t_d$ from $t_L=0$ (\autoref{fig:4_Int_stages}), in order to initiate droplet atomisation. The flow imposed by the blast wave results in the atomisation of the droplet. The timescale of the blast wave interaction is indicated by '$t$', where $t=0$ corresponds to the instant of wire-explosion. As shown in the velocity vs time plot at the top of \autoref{fig:4_Int_stages}, the blast wave reaches the droplet location after a time of $t=t_s$ from wire-explosion. The plot shows the schematic of the temporal variation of the velocity as imposed by the shock flow at the droplet location.
 
The interaction of the droplet with the shock flow generated by the shock-generation setup occurs in two stages, as established in previous experiments involving the same apparatus \citep{vadlamudi_insights_2024,vadlamudi2024Nano,aravind2025responsea,aravind2025responseb}. As shown in \autoref{fig:4_Int_stages}, the propagating blast wave initially imposes a velocity discontinuity at the droplet location at $t=t_s$, followed by a continuously decaying velocity profile ($\rm v_s$), which approaches zero around $t \sim t_{amb}$. In addition, a characteristic induced flow exits the shock tube in the form of a compressible vortex ring, which subsequently interacts with the droplet, resulting in droplet disintegration. The shock flow timescale can be normalised using $t_{amb}$ as $\tau \sim t/t_{amb}$, such that the decaying velocity profile behind the blast wave ($\rm v_s$) only occurs for $\tau<1$ and blast wave velocity profile becomes zero for $\tau>1$ (see \autoref{fig:4_Int_stages}). The blast wave velocity ($\rm v_s$) decay occurs at significantly faster timescales of $\sim \mathcal{O}(10^{-1})ms$, while the maximum velocity scales associated with $\rm v_s$ is of order $\rm v_{s,max} \sim 50-100$ m/s. On the other hand, the compressible vortex ring (CVR) travels at velocity scales of $\rm v_{CVR,prop} \sim 30-120$ m/s and arrives at the droplet location at $t=t_{CVR}$ (see \autoref{fig:4_Int_stages}). The interaction timescale of CVR with the droplet relatively slower compared to $\rm v_s$ interaction and is of the order of $\sim \mathcal{O}(10^0)ms$. The CVR arrives at droplet location at $t=t_{CVR}$ as shown in \autoref{fig:4_Int_stages}(top). This two-stage interaction of the shock flow with the droplet is depicted using green background ($\rm v_s$ interaction) and red background ($\rm v_{CVR}$ interaction), respectively. The yellow gradient region (in-between) represents the overlap where the CVR interaction has commenced ($t>t_{CVR}$), but $\rm v_s$ has not yet fully decayed to zero ($\tau<1$).

   \begin{figure}
    \centering
    \includegraphics[width=1\textwidth]{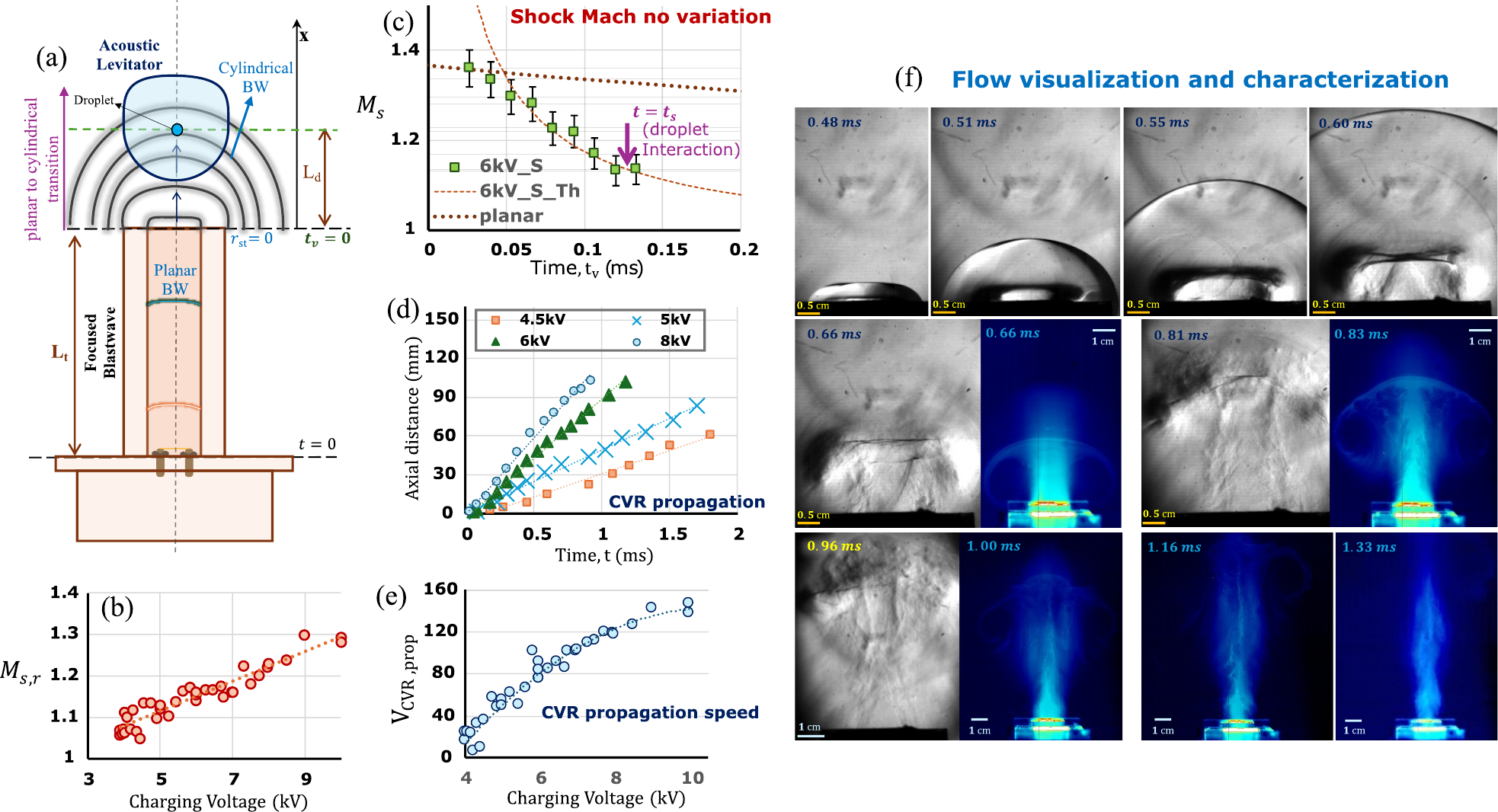}
    \caption{(a) Schematic of blast wave propagation, (b) $M_{s,r}$ vs Charging Voltage (kV), (c) Blast wave propagation: Theoretical vs Experimental, (d) CVR propagation for different charging Voltages, (e) CVR propagation speed variation ($\rm v_{CVR,prop}$), (f) Flow Visualization and characterization}
    \label{fig:5_FlowChar}
\end{figure}

\autoref{fig:5_FlowChar}a illustrates the schematic of blast wave propagation in the current experimental setup, showing an initially planar blast wave that transitions into a cylindrical blast wave upon exiting the shock tube. The same behaviour was observed in our earlier work \citep{vadlamudi_insights_2024}, where high-speed Schlieren imaging and PIV were used to visualize the flow and blast wave, using the same shock-generation setup, as shown in \autoref{fig:5_FlowChar}f. The flow velocity imposed by the blast wave at the droplet location can be obtained following the similar methodology as our previous experiments \citep{vadlamudi_insights_2024}.

A semi-analytical model by \cite{bach1970analytical} (which is valid for the $M_s$ range of current experiments i.e., $1.1 < M_{s,r} < 1.6$), a power-law density profile can be assumed behind the blast wave. Thus, for unsteady 1D adiabatic motion of perfect gas behind the expanding blast wave, density profile is obtained as:
\begin{equation}\label{Eq6: Shock}
   \psi\left(\xi,\eta\right)=\psi(1,\eta)\xi^{q(\eta)}
\end{equation}
whose exponent $q$ is obtained to be:
\begin{equation}\label{Eq8: Shock}   
q\left(\eta\right)=\left(j+1\right)\left[\psi\left(1,\eta\right)-1\right]
\end{equation}
using the mass integral obtained from the conservation of mass in the region enclosed by blast wave \citep{vadlamudi_insights_2024,bach1970analytical}. This has been elaborated in the supplementary material. 

Here, $\phi$ and $\xi$ are the non-dimensional parameters for velocity and density, as shown below:
\begin{equation}\label{Eq1: Shock}
   \phi\left(\xi,\eta\right)=\frac{u(r,t)}{{\dot{R}}_s(t)}
\end{equation}

\begin{equation}\label{Eq2: Shock}
   \psi\left(\xi,\eta\right)=\frac{\rho(r,t)}{\rho_o}
\end{equation}

where, $\theta\left(\eta\right)=\frac{R_s{\ddot{R}}_s}{{\dot{R}}_s},\ \xi=\frac{r}{R_s(t)},\eta=\frac{{c_o}^2}{{{\dot{R}}_s}^2}=\frac{1}{{M_s}^2},y\left(\eta\right)=\left(R_s/R_o\right)^{j+1}$  

$R_o$ is the characteristic explosion length, $R_s$ is the instantaneous shock radius, $u$ is the velocity field, $\rho$ is the density field and $c_o$ is the speed of sound. The boundary conditions at the shock front $\xi = 1$ are obtained from the standard normal shock relationship given by 
\begin{equation}\label{Eq4: Shock}
   \phi\left(1,\eta\right)=\frac{2(1-\eta)}{\gamma+1}
\end{equation}

\begin{equation}\label{Eq5: Shock}
   \psi\left(1,\eta\right)=\frac{\gamma+1}{\gamma-1+2\eta}
\end{equation}

Substituting the boundary condition at the origin as $\phi\left(0,\eta\right)=0$ in the mass conservation equation, the velocity profile is obtained as:
\begin{equation}\label{Eq10: Shock}   
   \phi=\phi(1,\eta)\xi\left(1-\Theta\ ln\xi\right)
\end{equation}
where, $\Theta=\frac{-2\theta\eta}{\phi(1,\eta)\psi(1,\eta)}\frac{\partial\psi\left(1,\eta\right)}{\partial\eta}$

After obtaining the $\theta$ vs $\eta$ relation following a similar methodology as shown by \cite{bach1970analytical}, Eq. \ref{Eq10: Shock} can be used to obtain the temporally decaying velocity profile ($\rm v_s$) at a given location ‘r’ behind the blast wave for a given $M_s$. 

The shock trajectory has been experimentally obtained by experimentally tracking the blast wave location in high-speed Schlieren (along the centreline) and this enables the estimation of Mach number ($M_s$) of the propagating blast wave at different instances. The data points shown in \autoref{fig:5_FlowChar}c plot shows the temporal variation of Blast wave Mach number ($M_s$) for a charging voltage of 6kV. The blast wave model has been theoretically validated \citep{bach1970analytical} by estimating the shock trajectory using the following expression and comparing with experimental trajectory.
\begin{equation}\label{Eq11: Shock}   
   \frac{c_ot}{R_o}=-\frac{1}{2}\int_{0}^{\eta}\frac{y^\frac{1}{j+1}d\eta}{\theta\eta^{1/2}}
\end{equation}
where, $\theta(\eta)$ is shock decay coefficient and $y\left(\eta\right)=\left(R_s/R_o\right)^{j+1}$ is dimensionless instantaneous shock radius.
 
Thus, the theoretical shock trajectory is estimated for a planar blast wave (inside shock tube) originating at the copper-wire (connected to electrodes) at $t=0$, as shown in \autoref{fig:5_FlowChar}a. It has been established in previous work that the blast wave transitions to cylindrical blast wave with its centre of curvature at the shock tube exit. Thus, the shock trajectory outside the shock tube has been obtained by considering the blast wave exiting shock tube (at $t_v=0$) having a cylindrical geometry with its origin at shock tube exit (distance $L_d$ from copper-wire), i.e., $r_{st}=0$. These have been plotted in \autoref{fig:5_FlowChar}c, where the brown dotted line and orange dashed line represent theoretical shock trajectories for planar and cylindrical blast waves, respectively, which show good agreement with the experimental trajectory. The pink arrow shows the instance of blast wave interaction with the droplet, i.e., the time instant ($t=t_s$) at which the blast wave reaches the droplet location (6.5cm from shock tube exit). The Mach number value when the blast wave reaches the droplet is considered as the reference Mach number ($M_{s,r}$) for the corresponding shock strength or charging voltage (kV). The variation of $M_{s,r}$ with charging voltage (kV) has been shown in \autoref{fig:5_FlowChar}b. 

As explained before, a compressible vortex ring (CVR) has been observed to follow the blast wave as shown in the simultaneous high-speed Schlieren and Mie-scattering imaging (see \autoref{fig:5_FlowChar}f). The variation of the vortex propagation speed with charging voltage (kV) is plotted in \autoref{fig:5_FlowChar}e.

It is to be noted that the major difference from our previous blast wave interaction experiments is that the droplet is placed at 3.5 cm from the shock-tube exit in previous experiments, compared to 6 cm in the current experiments. This has resulted in a higher decay of the blast wave strength at the droplet location in current experiments in comparison for the same shock tube setting. However, as the CVR propagates with nearly constant velocity (see \autoref{fig:5_FlowChar}d), the CVR velocity scales are similar in previous and current experiments for a given charging voltage (kV).

\FloatBarrier
\subsection{Global Observations of Droplet Atomization}\label{subsec:global Obs} \addvspace{10pt}

\begin{figure}
    \centering
    \includegraphics[width=1\textwidth]{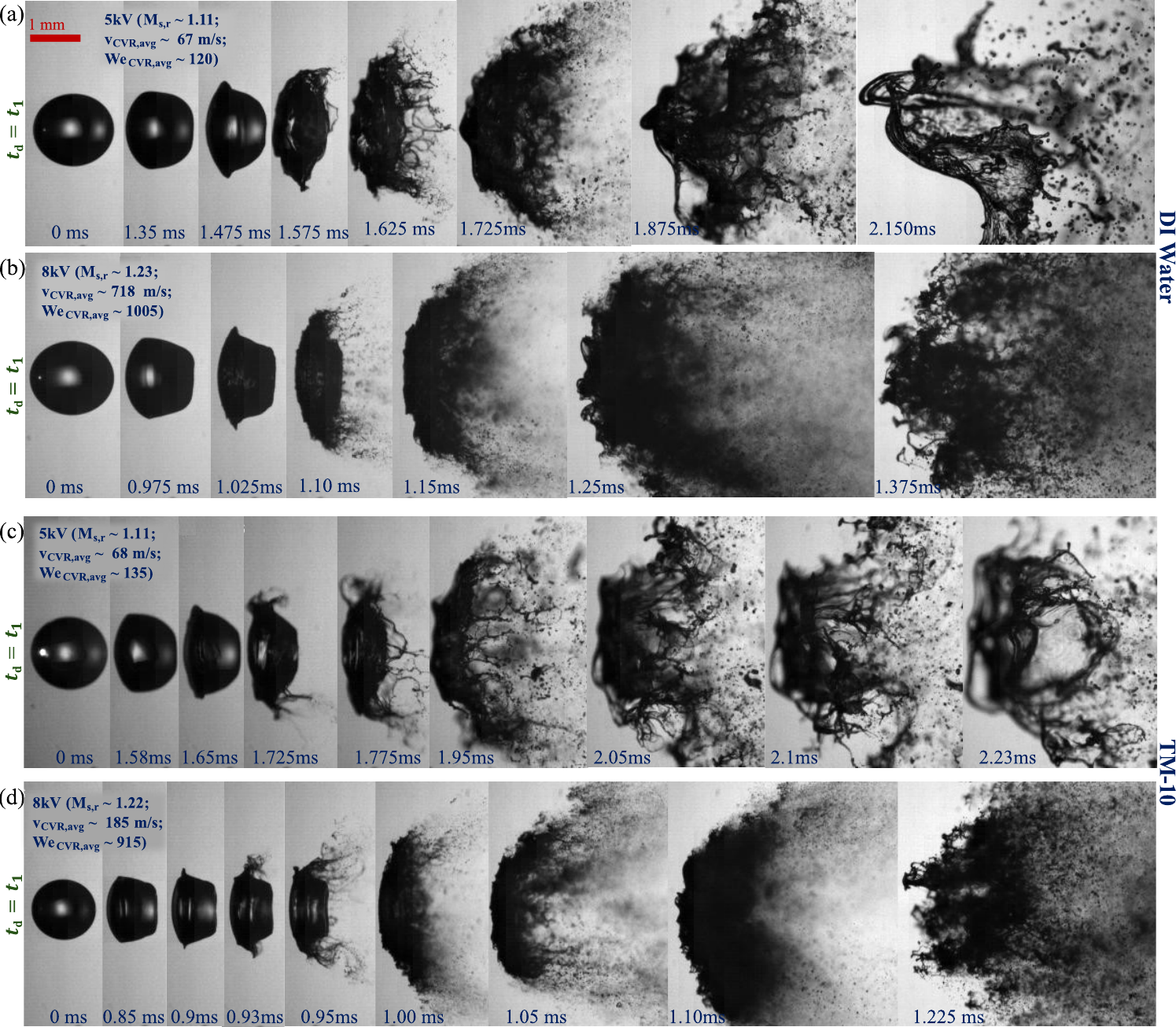}
    \caption{Timeseries snapshots of atomization in (a,b) DI water (\textbf{Movie 1}) and (c,d) TM-10 nanofluid (\textbf{Movie 3}) at two charging voltages.}
    \label{fig:6_GlobalObs}
\end{figure}

\autoref{fig:6_GlobalObs} shows the timeseries snapshots of droplet atomization at two different shock strengths ($M_{s,r}\sim 1.1 \&  \ 1.23$), for cases with no shell formation - namely, pure DI water (\autoref{fig:6_GlobalObs}a,b) and TM-10 nanofluid (\autoref{fig:6_GlobalObs}c,d) droplets, at low shock-delay ($t_d = t_1$). The two shock strengths $M_{s,r}\sim 1.1$ (\autoref{fig:6_GlobalObs}a,c) and $M_{s,r}\sim1.23$ (\autoref{fig:6_GlobalObs}b,d) correspond to 5kV and 8kV charging voltages, respectively. As the shock delay ($t_d$) for all the cases shown in \autoref{fig:6_GlobalObs} is $t_d\sim t_1$, they correspond to the early stages of evaporation i.e., steady evaporation regime (after the initial pre-heating time, see \autoref{fig:1_stagesOfEvap}a). Thus, the droplet surface temperature for these is at wet-bulb temperature. The corresponding $M_{s,r}$, velocity imposed by CVR ($\rm v_{CVR,avg}$), and the Weber numbers ($We_{CVR,avg}$) (based on $\rm v_{CVR}$, instantaneous droplet diameter during interaction, $D$ and surface tension at wet-bulb temperature) have been shown in \autoref{fig:6_GlobalObs}. 

\begin{figure}
    \centering
    \includegraphics[width=1\textwidth]{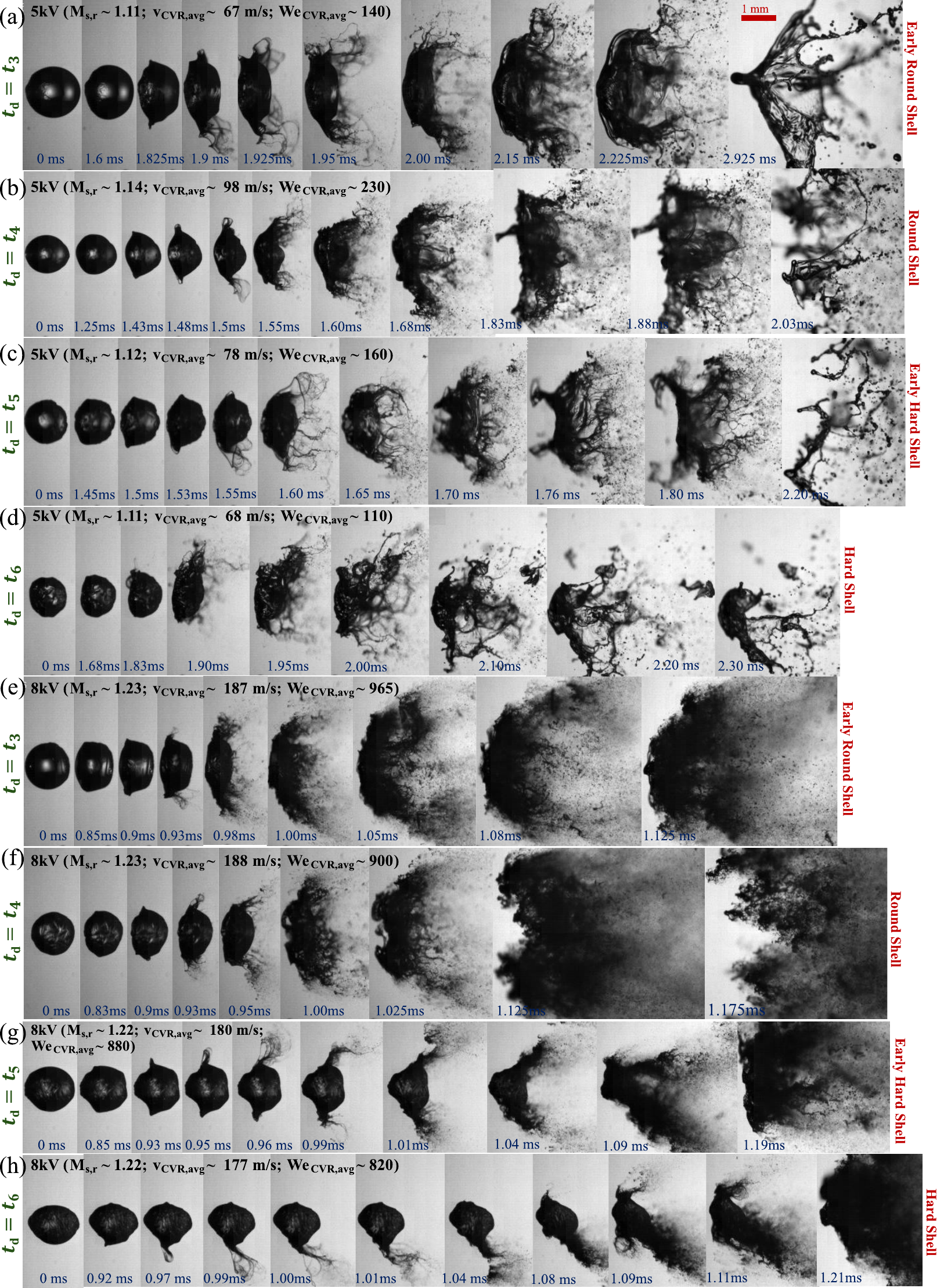}
    \caption{Droplet atomization timeseries for different shock delays - $t_L=t_d \simeq t_3$ to $t_6$, for two charging voltages: (a-d) 5kV and (e-h) 8kV.}
    \label{fig7:GlobalObv_TM10Delay}
\end{figure}

In \autoref{fig:6_GlobalObs}, since the Weber number ($We_{CVR,avg}$) is $ > 100$, all cases exhibited shear-induced entrainment (SIE) as the dominant breakup mode \citep{pilch1987use}. Experiments showed the SIE: sheet-stripping mode of breakup with KH-waves modulating the sheet formed from windward face. As shown in \autoref{fig:6_GlobalObs}, a sheet forms from the liquid on the windward side of the droplet due to aerodynamic shear. As this sheet extends outward, it undergoes secondary atomisation through the formation of ligaments. For higher shock strengths (8kV), the cascading effect of the KH-waves and the SIE or shear-stripping is more pronounced due to high Weber number ($We_{CVR,avg}\sim 1000$), as shown in \autoref{fig:6_GlobalObs}d. On the other hand, the lower shock strengths (5kV) tend to exhibit multi-mode breakup, specifically the bag-on-sheet mode, where Rayleigh–Taylor (RT) instabilities trigger the formation of multiple smaller bags within the shock-induced sheets (SIE), as shown in \autoref{fig:6_GlobalObs}c at 1.7ms. This behaviour corresponds to an intermediate Weber number range ($100 < We_{CVR,avg} < 1000$). At lower shock strengths, atomization was incomplete, and portions of the liquid remained intact in the form of liquid sheets following the interaction (\autoref{fig:6_GlobalObs}a,c beyond 2.15ms). Additionally, Rayleigh-Taylor piercing, RTP (i.e., bag formation) is also observed for lower shock strengths during later stages of interaction as seen in \autoref{fig:6_GlobalObs}c at 2.05ms and 2.23ms. It is to be noted that the primary difference between $t_d = t_1$ (heated droplet in the steady evaporation regime) and $t_d = t_0$ (non-heated droplet) is the droplet temperature. The elevated temperature in the heated droplet leads to reduced surface tension and thus a little higher Weber number compared to the non-heated case. 

The \autoref{fig7:GlobalObv_TM10Delay} shows the timeseries of the nanofluid droplet for higher shock delays ($t_d>t_1$), which fall in the unsteady evaporation regime where the effects of porous shell formation start to become noticeable. As mentioned before, the shock delays $t_d=t_3\ \& \ t_4$ correspond to the gelatinous shell regime, and similarly, $t_d=t_5 \ \& \ t_6$ correspond to the hard-solid shell regime. \autoref{fig7:GlobalObv_TM10Delay}(a-d) show the timeseries of the TM-10 nanofluid droplet atomisation during interaction with lower shock strengths ($M_{s,r}\sim 1.12$) for various shock delays ($t_d\sim t_3$ to $t_6$). Similarly, \autoref{fig7:GlobalObv_TM10Delay}(e-h) correspond to the higher shock strength cases ($M_{s,r}\sim 1.23$). 

\subsubsection{Gelatinous shell regime}
It can be evidently observed from \autoref{fig7:GlobalObv_TM10Delay}a,b,e,f that the droplet before interaction exhibits surface corrugations, deformations, and undulations, which are indicative of the shell formation on the surface of the nanofluid droplet, corresponding to $t_d=t_3 \ \& \ t_4$ (gelatinous shell regime). Near the droplet surface, convective and diffusive transport lead to the preferential accumulation of nanoparticles. As the local volume fraction approaches a critical threshold, interparticle interactions (such as van der Waals forces, hydrogen bonding, agglomeration etc.) promote the formation of a permeable, percolating network \citep{le2024sol}. This loosely connected network of colloidal silica and aggregates, containing trapped interstitial water, gives rise to a highly viscous, gelatinous (soft-solid) shell at the droplet surface. This effect becomes more pronounced during later stages of the gelatinous shell regime, i.e., \autoref{fig7:GlobalObv_TM10Delay}b,f, compared to the early stage of the gelatinous round shell regime (see \autoref{fig7:GlobalObv_TM10Delay}a,e).

\begin{figure}
    \centering
    \includegraphics[width=0.75\textwidth]{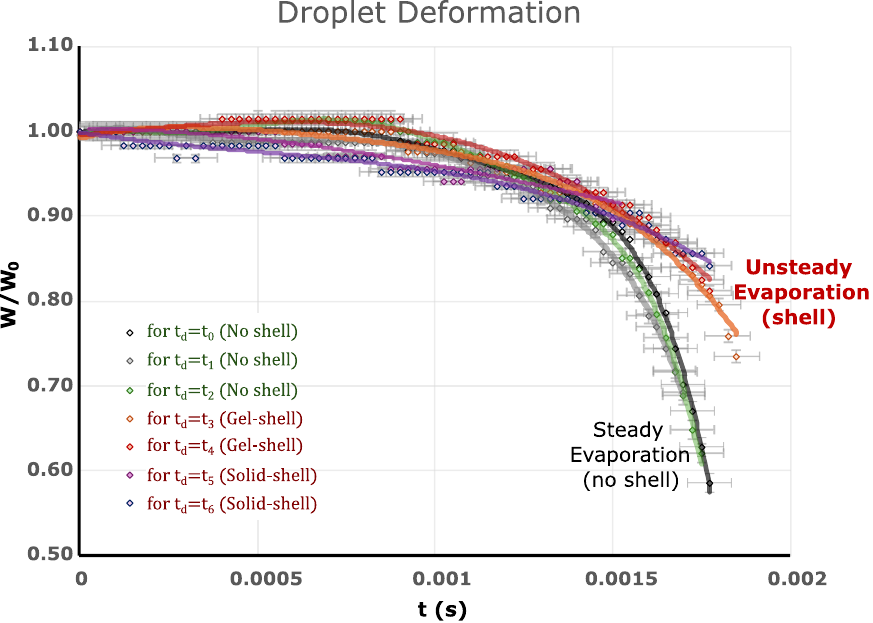}
    \caption{Droplet Deformation rate (Normalised streamwise width of the droplet with respect to initial value): Comparison for different delays ($t_d$) for the same shock strength ($M_{s,r}\sim1.12$), showing deformation rate for no-shell ($t_0,t_1,t_2$) and shell-formation ($t_3,..t_6$) cases respectively.}
    \label{fig7p5:DropletDef_TM10_Del}
\end{figure}

In this regime, during shock flow interaction with the droplet, it can be seen that the droplet showed significantly higher resistance to deformation compared to lower shock delay ($t_d < t_2$). This is evident from \autoref{fig7p5:DropletDef_TM10_Del}, where droplet deformation is significantly slower during unsteady evaporation regime ($t_d = t_3$ to $t_6$) compared to lower delays. Interestingly, the deformation rate of TM10 droplet becomes marginally faster from no-heating case $t_d=t_0$ (\textbf{Movie 2}) to $t_d=t_2$ (steady evaporation regime), which can be attributed to a slightly reduced surface tension and viscosity due to higher temperature. \autoref{fig7p5:DropletDef_TM10_Del} also shows a sharp reduction in the droplet deformation rate from the no-heating ($t_d=t_0$), steady evaporation ($t_d=t_1$) and transition ($t_d=t_2$) phase (\textbf{Movie 4}) to the post sol-gel transition ($t_d\ge t_3$), marking the onset of gelatinous surface shell formation. Unlike in the steady evaporation regime, the gelatinous shell regime exhibits distinct breakup dynamics. Following initial deformation, viscous liquid sheet protrusions emerge from the equatorial plane of the droplet (see \autoref{fig7:GlobalObv_TM10Delay}b at 1.48ms). As these sheets extrude outward, multiple smaller bags form in the equatorial sheet due to the modulation of Rayleigh–Taylor (RT) instabilities, characteristic of the bag-on-sheet mode of multimode breakup. Notably, this mode persists even under high shock strengths (8kV), as observed in \autoref{fig7:GlobalObv_TM10Delay}a,b,e,f. Furthermore, the bag-on-sheet breakup leads to the rupture of the viscous equatorial protrusion, which in turn disrupts the gelatinous shell enclosing the droplet. This rupture facilitates the sudden release of the inner liquid, previously confined within. The resulting collapse and disintegration of the gelatinous shell is followed by atomisation of the droplet: at lower shock strengths (5kV), this occurs via multibag formation driven by RT piercing, while at higher shock strengths (8kV), atomization proceeds through shear-induced entrainment (SIE) and shear stripping, as seen in \autoref{fig7:GlobalObv_TM10Delay}a,b,e,f.

\subsubsection{Solid shell regime}
As the shock delay ($t_d$) is further increased, the gelatinous shell undergoes additional agglomeration, leading to its hardening and solidification into a solid shell, as observed in \autoref{fig7:GlobalObv_TM10Delay}c,d,g,h. Since the TM-10 nanofluid is composed of silica nanoparticles ($\mathrm{SiO_2}$), the local silica concentration increases with time, resulting in a glassy appearance of the solidified shell at $t=0$ in \autoref{fig7:GlobalObv_TM10Delay}c,d,g,h. \cite{noppari2025analyzing} showed that after the sol-gel transition, the colloidal silica network becomes a dense, rigid structure with solid-like characteristics. 
In the solid-shell regime, the initial droplet deformation is significantly altered due to the presence of a solid shell which is partially or fully covering the droplet surface. Since the shell is composed of silica particles, it exhibits brittle behaviour, characterised by minimal deformation and cracking (after the shell is fully formed), as shown in \autoref{fig7:GlobalObv_TM10Delay}h. However, stochastic variations in the shell solidification process that arise from non-uniform deposition or drying lead to partial deformation before the shell-cracking, as observed in \autoref{fig7:GlobalObv_TM10Delay}c,d,g. After the solid shell gets punctured, the entrapped liquid rushes out in the form of a sheet, which undergoes RT piercing, resulting in a bag-on-sheet mode of multimode breakup, similar to the gelatinous shell regime. Once, the outer viscous layer is fully ruptured, the inner liquid rushes out in the form of a jet, which atomises similar to a jet in a crossflow, as shown in \autoref{fig7:GlobalObv_TM10Delay}d,g. The atomization mode of this liquid jet depends on the flow velocity imposed during the shock flow interaction. Once the shell ruptures at one location, it also starts to peel-off from other locations due to the loss of structural integrity, as shown in \autoref{fig7:GlobalObv_TM10Delay}g. In the early hard-shell stage, shown in \autoref{fig7:GlobalObv_TM10Delay}c,g, the solid shell is not yet fully developed and represents a transitional phase between the gelatinous and solid-shell regimes.

The initial shell failure is observed in the form of localised puncture and fragmentation at the equatorial region (\autoref{fig7:GlobalObv_TM10Delay}d at 1.90ms). A subsequent secondary rupture event (termed shell rupture-II) occurs near the forward stagnation point on the windward side, driven by shell delamination due to local shear near the location (see \autoref{fig7:GlobalObv_TM10Delay}c at 1.70ms). In the later stages, the shell undergoes catastrophic breakup on the leeward side, where it delaminates and peels off (see \autoref{fig7:GlobalObv_TM10Delay}d at 2.00ms). This is caused by the propagation of earlier cracks and a progressive loss of structural integrity in the brittle shell, resulting in complete disintegration of the droplet and shell.

\FloatBarrier

\begin{figure}
    \centering
    \includegraphics[width=1\textwidth]{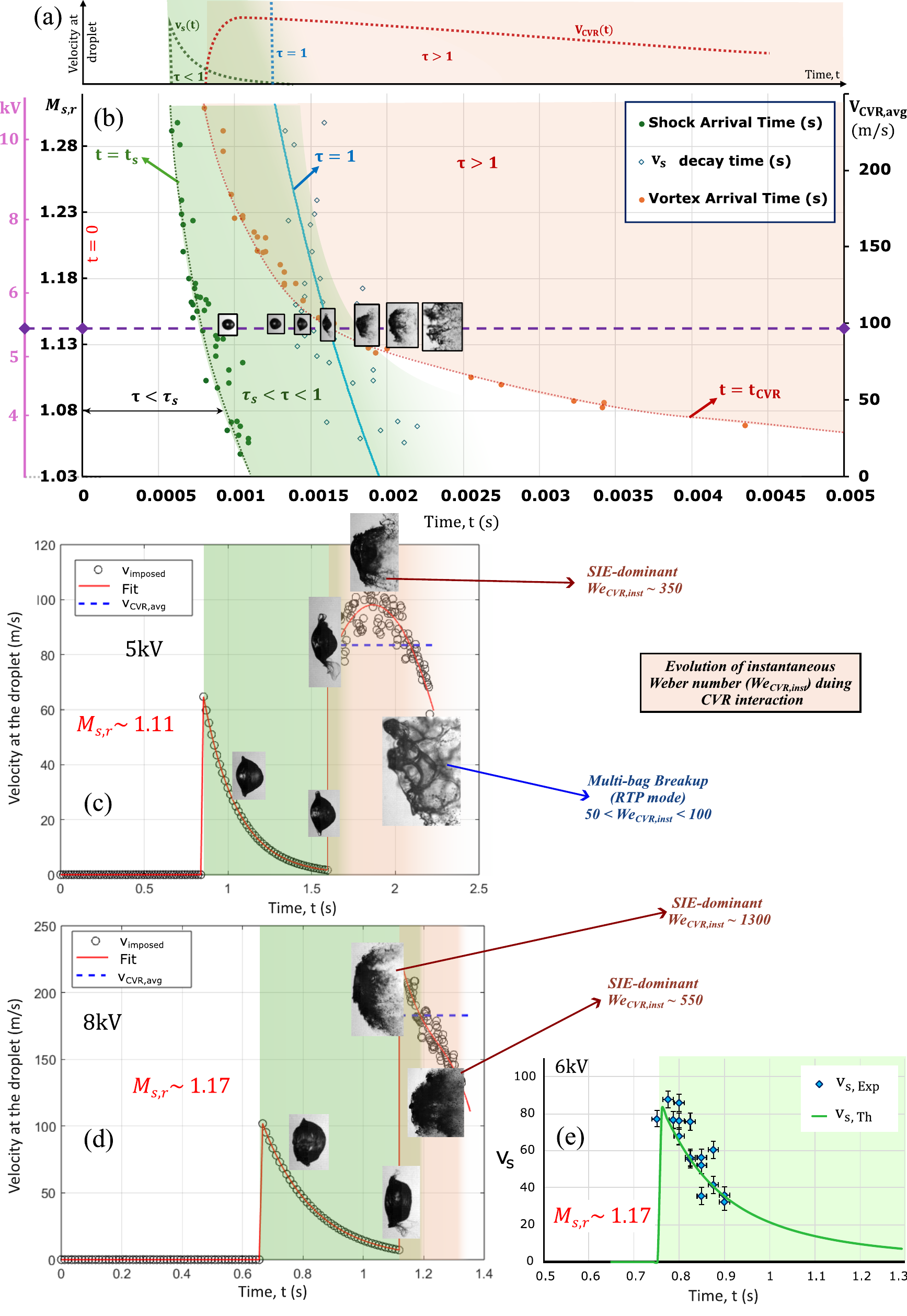}
    \caption{(a) Schematic of temporal variation of the imposed velocity at the droplet occurring at different timescales during interaction.}
    \label{fig8:timescales_Stages}
\end{figure}

\begin{figure}\ContinuedFloat
  \caption{(continued) (b) Timescales of $\rm v_s$ (green region) and $\rm v_{CVR}$ (orange region) interaction with the droplet plotted with the y-axis representing the reference shock Mach number ($M_{s,r}$) along with the corresponding charging voltage in kV (pink axis). The secondary axis represents the average velocity of CVR (in $m/s$). Green and orange data points represent the shock ($t_s$) and CVR arrival time ($t_{CVR}$) from the wire-blast ($t=0$). The solid blue line represents the time instant where $\rm v_s \rightarrow 0$ ($\tau=1$). (c,d) The temporal variation of the velocity imposed (at the droplet location), $\rm v_{imposed}$, during interaction for different $M_{s,r}$, plotted using circular data points. The blue horizontal dotted line represents the average $\rm v_{CVR}$ ($\rm v_{CVR,avg}$). The green and orange backgrounds represent $\rm v_s$ and $\rm v_{CVR}$ interaction with the droplet, respectively. (e) Temporal variation of $\rm v_s$ at the droplet: theoretical vs experimental. Sequential droplet images are displayed as sub-figures to illustrate the representative stages of the atomisation process occurring at different timescales.}
\end{figure}

\subsection{Timescales of shock flow interaction and droplet atomization}\label{sec:Int_timescales}
The time scale of the atomization depends on the time scale of the two stage interaction of the shock-flow. As explained, the first stage interaction with the decaying velocity ($\rm v_s(t)$) of the blast wave is short-lived ($\sim \mathcal{O}(10^{-1})ms$) and the second stage interaction with CVR ($\rm v_{CVR}(t)$) occurring over a timescale of $\sim \mathcal{O}(10^0)ms$. The interaction with $\rm v_s$ only results in droplet deformation, while the droplet atomisation actually occurs only after the interaction of $\rm v_{CVR}(t)$. The droplet atomisation was only observed to occur after the CVR reaches the droplet, and this observation has been consistent with the previous experiments \citep{vadlamudi_insights_2024,vadlamudi2024Nano,sharma_shock-induced_2023,chandra_shock-induced_2023,rao2025secondary}. 

The same has been depicted in the schematic shown in \autoref{fig8:timescales_Stages}a showing the two stages of interaction: one with decaying velocity behind blast wave, $\rm v_s$ (represented using green background) and the other with CVR (represented using orange background). The green region represents $\tau <1$ regime which indicates non-zero $\rm v_s$ and the orange region corresponds to $\tau>1$, where $\rm v_s$ has fully decayed to zero, as shown in \autoref{fig8:timescales_Stages}a). The end of the green region, indicated by vertical blue line indicates $\tau=1$ where, $\rm v_s \rightarrow 0$. The theoretical blast wave model discussed in \autoref{sec:Flow_field} is used to obtain the velocity variation due to the blast wave ($\rm v_s$) based on the variation of the instantaneous Mach number ($M_s$) at the droplet location. This has been shown to be in good agreement with the theoretical velocity, $\rm v_s$, as shown in \autoref{fig8:timescales_Stages}e, which is consistent with the previous experiments \citep{vadlamudi_insights_2024,vadlamudi2024Nano}. The plot in \autoref{fig8:timescales_Stages}b illustrates the variation of $M_{s,r}$ and the average CVR velocity ($\rm v_{CVR,avg}$) for different charging voltages (i.e., shock strengths), with time $t$ on the x-axis ($t = 0$ corresponds to the wire explosion). The green data points represent $M_{s,r}$ plotted against the shock arrival time ($t_s$, in seconds), with the corresponding charging voltage (in kV) indicated on a pink axis. The orange data points show the CVR arrival time ($t_{CVR}$) along with the associated $\rm v_{CVR,avg}$ (in m/s), plotted on the secondary axis (right). All axes are aligned such that any horizontal line (highlighted by a purple dashed line) corresponds to a specific experimental case, with its associated values of $M_{s,r}$, charging voltage, and $\rm v_{CVR,avg}$ of that specific case. 

It can be observed from \autoref{fig8:timescales_Stages}a that from left to right (with increase in time from $t=0$, i.e., wire explosion), initially the droplet experiences no velocity for $t<t_s$. After the blast wave passes past the droplet at $t=t_s$ (green data points), $\rm v_s$ is imposed on the droplet in green region, which decays to zero at $\tau = 1$ (indicated by blue solid line and points). Subsequently, when the CVR reaches the droplet at $t=t_{CVR}$ (orange data points), $\rm v_{CVR}$ is imposed on the droplet in orange region. It can be observed that shock and CVR arrival times ($t_s$ and $t_{CVR}$) decrease with an increase in shock strength due to an increase in velocity scales (from bottom to top). Furthermore, at lower shock strength, CVR arrival time at the droplet is observed to be significantly delayed due to the low velocity scales of CVR, resulting in a wait time between decay of $\rm v_s$ (i.e., $\tau=1$) and CVR arrival $t_{CVR}$. This has also been observed experimentally, where the droplet atomisation and disintegration are delayed for low shock strengths. As the shock strength is increased, $t_{CVR}$ decreases, resulting in faster arrival of CVR at the droplet, thus shortening the atomisation timescale. When the shock strength is increased further, $t_{CVR}$ becomes less than $t_{amb}$ ($\rm v_s \rightarrow 0$ time, i.e., $\tau=1$), resulting in an overlap zone (between green and orange regions), where CVR interaction occurs even before $\rm v_s$ fully decays to zero. A sample sequence of experimental images capturing the droplet atomisation process is shown as sub-figures, along the purple dashed horizontal line, depicting the interaction at different timescales.

In addition to the theoretical estimate of the temporal variation of $\rm v_{s,Th}$ obtained using \autoref{sec:Flow_field}, it is also necessary to evaluate the local velocity variations at the droplet velocity during CVR interaction. Although the PIV measurements shown in \autoref{fig:5_FlowChar} provide an estimate of the velocity scale imposed by the CVR, their temporal resolution is limited by the 3000 fps acquisition rate in double-frame mode. Consequently, the current experiments require higher temporal and spatial resolution to capture the velocity variations more accurately. Thus, to pre-seed the CVR, DEHS (Di-Ethyl-Hexyl-Sebacate) particles were introduced into the shock tube prior to the wire explosion and high-speed shadowgraph imaging was then performed (without the droplet) to visualise the flow. Background subtraction was applied to enhance contrast, and a particle tracking algorithm was used to track the seeder particles as the CVR passed through the camera's region of interest (ROI), corresponding to the droplet location. From this analysis, a matrix containing the time instant, axial position, and velocity of each tracked particle was generated. It is important to note that the transverse velocity variation was minimal, given that the CVR size is much larger than the droplet diameter. Hence, the transverse variation had been neglected for simplicity, and the axial velocity data were then interpolated over time and space to obtain a continuous velocity field (see Supplementary Figure S5(b,c)).

Furthermore, it has been observed from experiments (\autoref{fig:6_GlobalObs}) that during the droplet interaction and atomization, the droplet is also pushed downstream. Thus, the center-of-mass of the droplet has been tracked to obtain the time-varying location of the moving droplet for different shock strengths. Thus, the accurate estimate of the velocity imposed by CVR ($\rm v_{CVR}(t)$) at the droplet location has been obtained using the velocity field (see Supplementary Figure S5(b,c)) by incorporating the droplet movement. Finally, the temporal variation of the velocity imposed at the droplet location ($\rm v_{imposed}$) during different stages of interaction has been obtained using $\rm v_{s,Th}$ (for $t<t_{CVR}$) and $\rm v_{CVR}(t)$ (for $t>t_{CVR}$):

\begin{equation}
    \rm v_{imposed}(t) = 
    \begin{cases}
    \rm v_{s,Th}(t), & \text{for } t < t_{CVR}\\
    \rm v_{CVR}(t), & \text{for } t > t_{CVR}
    \end{cases} 
    \label{Eq:1 vimposed}
\end{equation}

where, $\rm v_{s,Th}$ is the temporally decaying velocity imposed by blast wave (obtained theoretically), and $\rm v_{CVR}(t)$ is the temporally varying velocity imposed at the droplet location during CVR interaction (obtained experimentally). Thus, $\rm v_{imposed}(t)$ obtained from \autoref{Eq:1 vimposed} has been plotted in \autoref{fig8:timescales_Stages}c,d for different shock strengths, showing the two-stages of interaction with $\rm v_s$ and $\rm v_{CVR}$ with green and orange background, respectively. The sub-figures in \autoref{fig8:timescales_Stages}c,d show representative frames across the breakup sequence. After the initial deformation during $\mathrm{v_s}$ interaction, the subsequent $\mathrm{v_{CVR}}$ phase drives sheet formation and breakup. At the onset of CVR interaction ($\tau \rightarrow 1^+$), the velocity imposed at the droplet location reaches a local maximum and then decays.Owing to the temporal variation of velocity at the droplet location, the instantaneous Weber number also evolves in time. The instantaneous Weber number $We_{CVR}(t)$, computed from the CVR-imposed velocity at the droplet centroid (Supplementary Figure S5c), tries to capture the most relevant velocity scales driving breakup.

For low shock strength i.e., 5kV case ($We_{CVR,\mathrm{avg}} \sim 100$–$200$; \autoref{fig8:timescales_Stages}c), the instantaneous Weber number rises to $\sim 300$ near $\tau \rightarrow 1^+$ and later decreases below 100, to as low as $\sim 50$. Bags-on-sheet breakup appears early, with an equatorial sheet formed predominantly by shear, which is consistent with the RTP-SIE transition region (reported around $We \sim \mathcal{O}(10^2$–$10^3)$ in shock/impulsive flows \citep{sharma_shock_2021}). As the interaction proceeds and the instantaneous $We$ drops to a value as low as $\sim 50$, Rayleigh–Taylor piercing becomes dominant, producing multibag breakup. This is in qualitative agreement with observations with the literature \citep{sharma_shock-induced_2023}. At higher shock strength (8 kV; $We_{CVR,\mathrm{avg}} \sim 900$–$1000$), the instantaneous Weber number decreases from $\sim 1200$ at $\tau \rightarrow 1^+$ to $\sim 500$ at later times, and shear-induced stripping (SIE) persists throughout the atomisation sequence, consistent with the literature. This analysis examines how a time-varying Weber number ($We$) influences the evolution of atomisation characteristics, an aspect not explicitly addressed in previous studies on droplet aerobreakup \citep{sharma_shock-induced_2023}.


\FloatBarrier

\subsection{Temporal evolution of droplet composition, internal properties, and phase transition regimes}\label{sec:composition}

\begin{figure}
    \centering
    \includegraphics[width=1\textwidth]{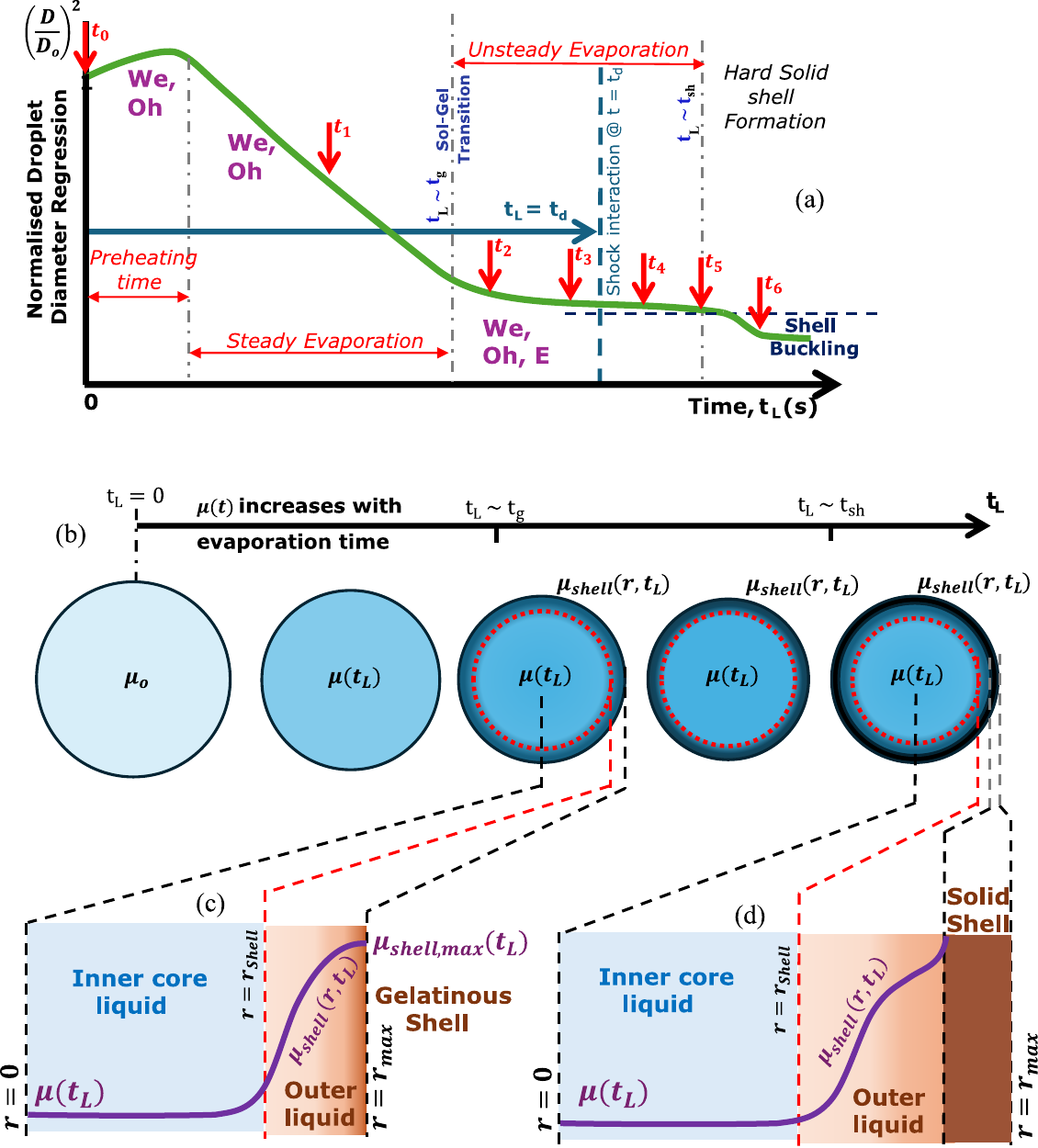}
    \caption{(a) Diameter regression plot showing different stages of evaporation in nanofluid droplet and different shock-delays ($t_d$) considered (see \autoref{tab:shock_Delays}), marked with red arrows. (b) The Schematic of the evolution of the droplet-interior, showing agglomeration, sol-gel transition, and time-dependent variation in properties inside the droplet.(c,d) Schematic of the radial variation of viscosity in inner core and outer liquid layer in (c) gelatinous shell and (d) Solid shell regimes. The different parameters that are relevant in each of the regimes is mentioned in purple font.}
    \label{fig9:Sol-Gel transition}
\end{figure}

Thus far, only the aerodynamic aspects of droplet atomisation have been discussed in detail; however, the TM-10 nanofluid droplet also undergoes compositional changes and shell formation, both of which significantly modify the breakup dynamics. \autoref{fig9:Sol-Gel transition}a shows the schematic of droplet diameter regression during nanofluid droplet evaporation. It depicts an initial preheating time, followed by the steady evaporation regime (linear regression of the square of droplet diameter) and finally the unsteady evaporation regime, where the droplet evaporation rate slows down due to porous gelatinous shell formation at the droplet surface, as discussed in \autoref{sec:sample_Char}. The sol-gel transition is observed to occur at $t_L=t_g$, where the particle aggregation reaches a critical packing fraction leading to gelation, forming a gelatinous shell near the droplet surface. As shown in \autoref{fig9:Sol-Gel transition}a, the key distinction at $t=t_0$ is that it represents the unheated baseline case. As heating begins from $t=t_0$, the droplet undergoes an initial preheating phase during which its temperature gradually rises from the ambient temperature ($T_{\inf}$) to the wet-bulb temperature $T_{wb}$. Once $T_{wb}$ is reached, the surface temperature of the droplet stabilises due to a dynamic balance between evaporative cooling and external heat input, thereby establishing a steady evaporation regime. During the preheating phase, the increase in temperature leads to a reduction in droplet properties such as surface tension and viscosity. Thus, Weber number and Ohnesorge number are relevant parameters for droplet atomisation in the pre-heating regime. 

On the other hand, during the steady evaporation regime, the absorbed heat is primarily utilised as the latent heat of vaporisation, leading to a constant rate of phase change at the droplet surface. Consequently, the droplet undergoes a steady reduction in size due to the continuous evaporation of the liquid phase, as illustrated in \autoref{fig9:Sol-Gel transition}a. This progressive loss of solvent results in a temporal increase in the concentration of nanoparticles within the droplet. This increase in nanoparticle concentration leads to a corresponding rise in the viscosity ($\mu$) within the droplet \citep{poon2012measuring,chen2007rheological,meyer2016viscosity,mewis2012colloidal}. In addition, the rising particle concentration within the droplet gives rise to a secondary effect of enhanced particle agglomeration that is driven by increased inter-particle interactions \citep{lu2013colloidal,chen2007rheological,selvakumar2017effective,gaganpreet2015viscosity,meyer2016viscosity,duan2011viscosity,mewis2012colloidal,metin2014aggregation}. Thus, during the steady evaporation regime, the droplet diameter ($D$) continuously decreases, accompanied by a progressive increase in the internal viscosity ($\mu(t_L)$), see \autoref{fig9:Sol-Gel transition}b. Thus, during steady evaporation, the simultaneous decrease in droplet diameter ($D$) and evolution of viscosity make the Weber and Ohnesorge numbers particularly relevant to aerobreakup in this regime. However, the variation in internal viscosity becomes increasingly prominent over time and is comparatively less significant in the initial stages of steady evaporation. 

The semi-empirical relation developed by \cite{krieger1959mechanism} for estimating the viscosity of a nanofluid ($\mu$) over the full range of particle volume fractions, can be used to estimate $\mu(t_L)$:
\begin{equation}
    \frac{\mu(t_L)}{\mu_o}=\left [ 1- \frac{\phi(t_L)}{\phi_m}\right ]^{-[\eta]\phi_m}
    \label{Eq:KD-model}
\end{equation}
where $\mu(t_L)$ and $\phi(t_L)$ are the instantaneous viscosity and particle volume fraction of nanofluid during evaporation at a given time $t_L$ and $\mu_o$ is the initial viscosity of the nanofluid. $\phi_m$ represents the maximum packing fraction for the constituent particles in the nanofluid at a given time instant ($t_L$). [$\eta$] represents the intrinsic viscosity, whose value for hard spheres (NPs) is $\sim 2.5$ \citep{chen2007rheological}. $\phi$ can be evaluated using the relation:
\begin{equation}
    \phi(t_L) = \frac{\frac{w_o}{\rho_p}}{\frac{w_o}{\rho_p}+\frac{1-w_o}{\rho_l}\left(\frac{D(t_L)}{D_o}\right)^3}
\end{equation}
Where $w_o$ is initial mass fraction of NPs, $\rho_p$ and $\rho_l$ are the densities of the base liquid and NPs. However, this model does not include the effects of nanoparticle agglomeration as $\phi(t_L)$ increases during evaporation.

Primarily, the tendency of particle agglomeration increases with the increase in concentration, and the agglomeration progressively increases with time (see \autoref{fig:1_stagesOfEvap}c). The particle agglomeration results in the increased contribution of the constituent particles to the overall viscosity. Thus, the particle packing fraction $\phi$ in \autoref{Eq:KD-model} is replaced with $\phi_a$ (apparent volume fraction due to aggregation), as the aggregates formed due to agglomeration occupy more volume due to interstitial voids and fractal structure \citep{gaganpreet2015viscosity}. $\phi_a$ can be estimated from the aggregate size ($r_a$). It is known that during evaporation, the individual nanoparticles (of size $r_p$) in the nanofluid agglomerate to form aggregates (of size $r_a$) in the bulk liquid medium. Thus, the temporal growth of the aggregate size needs to be evaluated. Hence, assuming identical spherical nanoparticles initially, Smoluchowski's aggregation formulation \citep{russel1991colloidal,zaharia2014coagulation} based on Brownian motion-driven particle collisions can be used to estimate the rate of change of number concentration $N(t_L)$:
\begin{equation}
    \frac{dN}{dt_L}=-k_{agg}N^2
\end{equation}
Where $k_{agg}$ is the aggregation rate constant.

Integrating the above equation gives:
\begin{equation}
    N(t_L)=\frac{N_o}{1+k_{agg}N_o \ t}
    \label{Eq:N(t)}
\end{equation}

\begin{equation}
    N_o = \frac{\phi_o}{\frac{4}{3}\pi r_p^3}
\end{equation}
The aggregation can be either perikinetic (brownian motion-driven) or orthokinetic (shear-driven) in nature \citep{crittenden2012mwh,bratby1980coagulation}
.
The Perikinetic aggregation rate constant ($k_{agg,pr}$) is given by:
\begin{equation}
    k_{agg,pr}=\frac{8k_BT}{3\mu}
    \label{eq:peri}
\end{equation}
where, $k_B$ is Boltzman constant and $T$ is temperature in Kelvin. On the other hand, the orthokinetic aggregation rate constant ($k_{agg,or}$) is given by:
\begin{equation}
    k_{agg,or}=\frac{4}{3}\alpha \dot{\gamma} \ r_p^3 
    \label{eq:ortho}
\end{equation} 
where, $\dot{\gamma}$ is shear rate inside the droplet and $\alpha$ is the collision efficiency factor for the particles during agglomeration \citep{crittenden2012mwh,bratby1980coagulation}. The shear rate ($\dot{\gamma}$) is obtained as $\dot{\gamma}\sim u_{str,l}/\delta$. $\delta$ is the thickness of the acoustic boundary layer or Stokes layer and $u_{str,l}$ is the streaming velocity scale inside an acoustically levitating droplet, scaled as $u_{str,l}\sim U^2/(R\omega)$, where $U$ is the external acoustic velocity amplitude scale in the levitator \citep{hasegawa2020transport}. 

However, in the context of current experiments, for an evaporating nanofluid droplet, the mode aggregation characteristics needs to be evaluated. Thus, different types of Peclet numbers are evaluated as follows:

\begin{subequations} \label{Eq:Pe}
\begin{align}
Pe_{int} \sim \frac{u_{str,l} \ r_{max}}{\mathcal{D}} \label{Eq:Pe:int} \\
 Pe_{evap} \sim \frac{v_{evp} r_{max}}{\mathcal{D}} \label{Eq:Pe:evap} \\
Pe_{ortho} \sim \frac{\dot{\gamma} r_p^2}{\mathcal{D}} \label{Eq:Pe:orth}
\end{align}
\end{subequations}

where, $r_{max}$ levitated droplet of radius ($r_{max}\sim D/2$), $\mathcal{D}$ is the diffusion coefficient obtained using Stokes-Einstein relation, $\mathcal{D}={k_BT}/({6\pi \mu r_p})$. $v_{evp}$ is the velocity scale corresponding to the flux of evaporating surface, $v_{evp}=-({dr_{max}}/{dt})$. Based on \autoref{Eq:Pe}, the Peclet number based on internal circulation is found to be of the order of $Pe_{int}\sim \mathcal{O}(10^3-10^4)$, indicating rapid mixing within the droplet's inner core and supporting the assumption of a uniform bulk liquid in this region. Furthermore, the Peclet number based on the evaporation-driven surface regression is found to be of the order $Pe_{evp}\sim \mathcal{O}(10^2-10^3)$, suggesting the potential for surface accumulation and subsequent shell formation. In contrast, the shear-based Peclet number is obtained as $Pe_{ortho}\ll 1$ indicating that Brownian diffusion dominates over orthokinetic aggregation in the system. 

Thus, as illustrated in the schematic in \autoref{fig9:Sol-Gel transition}b, the bulk of the droplet remains uniform during evaporation, characterised by a uniform viscosity $\mu(t_L)$. Beyond the time $t_L>t_g$, where gelation initiates at the surface, a gelatinous shell begins to form due to the accumulation, crowding, and agglomeration of particles near the droplet interface, which is also supported by $Pe_{evp}$ (\autoref{Eq:Pe:evap}). This region is referred to as the ‘outer liquid’ (see \autoref{fig9:Sol-Gel transition}b), within which a steep radial increase in viscosity is observed from $r=r_{max}$ towards the droplet surface ($r=r_{max}$).

Gelation in the outer liquid layer occurs around $t_L\sim t_g$, when nanoparticle agglomeration leads to a local increase in particle concentration in the outer layer such that the local volume fraction $\phi_{outer}$ reaches a critical threshold, $\phi_g$. This marks the formation of a particle network, resulting in a significant rise in viscosity and a substantial reduction in fluidity within the outer layer. Furthermore, after the onset of gelation ($t_L>t_g$), the viscosity in the outer liquid layer, denoted by $\mu_{shell}(r_{max},t_L)$, evolves differently from the bulk liquid viscosity $\mu(t_L)$, as illustrated by the contrasting blue and brown regions in \autoref{fig9:Sol-Gel transition}c. At low concentrations, particles undergo Brownian motion and diffuse freely throughout the sample. As the concentration increases, the particles pack together randomly, causing a sharp rise in viscosity. \cite{weeks2017introduction} showed that below the glass transition concentration, Brownian motion still allows the sample to flow and equilibrate, so it behaves as a liquid. Above this threshold, however, equilibration is no longer possible on experimental time scales, and the sample develops a yield stress, behaving macroscopically like an elastic solid. In the present experiments, the outer layer likewise begins to act as an elastic membrane, exhibiting measurable elasticity \citep{di2012rheological} as the particle concentration rises rapidly, thereby resisting droplet deformation and disintegration. Consequently, during this regime, the Weber number, Ohnesorge number, and elastic modulus become important parameters governing aerobreakup, influenced respectively by droplet size regression, viscosity changes from agglomeration, and the evolving elastic properties of the shell (see \autoref{fig9:Sol-Gel transition}a).  

As evaporation progresses further, the local particle volume fraction in the outer shell region ($\phi_{outer}$) continues to increase, eventually approaching a limiting value $\phi_s\sim 0.64$, based on the assumption of random close packing for spherical NPs. This leads to the solidification of a thin layer at the edge of the outer liquid region, as shown in \autoref{fig9:Sol-Gel transition}d. Solidification occurs within the viscous outer layer (brown region) at the radial position ($r = r_{solid}$), where the local particle volume fraction ($\phi_{outer}$) increases sharply and approaches the maximum packing limit ($\phi_m\sim \phi_s$). Correspondingly, the viscosity exhibits a steep radial increase near ($r = r_{solid}$), beyond which the material ceases to behave as a liquid and begins to exhibit solid-like characteristics. As TM-10 nanofluid used in the current experiments contains silica nanoparticles, once solidification occurs, the resulting shell exhibits glassy, brittle behaviour, consistent with the literature \citep{di2012rheological,duffours1994weibull,woignier2015mechanical}. \cite{noppari2025analyzing, weeks2017introduction} showed that as the silica network becomes more crowded due to agglomeration at the liquid-air interface, the particles form a dense, rigid structure, having high viscosity and solid-like behaviour, which has been attributed to the exponential increase in mechanical strength observed with increasing silica content.

Building on this, the value of the maximum packing fraction $\phi_m$ used in the Krieger–Dougherty model used in \autoref{Eq:KD-model} can be reasonably approximated as $\phi_s$ in the case of spherical nanoparticles without agglomeration, due to their efficient packing in a random close-packed structure. However, as agglomeration progresses, the formation of irregular and loosely bound aggregates reduces the overall packing efficiency. Consequently, the effective maximum packing fraction $\phi_m$ and therefore $\phi_s$ decreases, reflecting the diminished efficiency of packing in case of agglomerated clusters compared to the monodisperse spherical nanoparticles, as shown by \citep{di2012rheological,chen2007rheological}. According to the characterisation study by \cite{di2012rheological} on LUDOX HS-40 silica suspensions, liquid-like Newtonian behaviour with minimal agglomeration was exhibited for $0.22 \le \phi \le 0.30$ (sol-state). For higher volume fractions ($0.31 \le \phi \le 0.35$), agglomeration had been reported. HS-40 exhibited elastic behaviour having a high elastic modulus ($G\sim \mathcal{O}(10^4-10^5)Pa$) for volume fractions of $0.35 \le \phi \le 0.47$, corresponding to the gel-state. Finally, it was reported to undergo a glass transition by solidifying into a brittle solid for a particle concentration of $\phi\sim 0.51$.

Therefore, since the orthokinetic Peclet number $Pe_{ortho}\ll1$ (\autoref{Eq:Pe:orth}), the brownian diffusion-dominant perikinetic agglomeration can be considered for estimating agglomeration. Hence, $k_{agg,pr}$ obtained using \autoref{eq:peri} can be used for the calculation of aggregate size evolution ($r_a(t_L)$). Thus, assuming fractal aggregation governed by the aggregation constant $k_{agg,pr}$ for the particle number evolution $N(t_L)$ (see \autoref{Eq:N(t)}), the temporal variation of the aggregate size ($r_a(t_L)$) can be evaluated as:
\begin{equation}
    r_a(t_L)=r_p\left[ \frac{N_o}{N(t_L)} \right]^{1/D_f}
    \label{Eq:r_a}
\end{equation}
where, $N_o$ is initial particle number concentration, $D_f$ is the fractal dimension and $r_p$ is the radius of the nanoparticle.

Thus, the modified \cite{krieger1959mechanism} model that accounts for agglomeration, for estimating the temporal nanofluid viscosity variation ($\mu(t_L)$) is given by:

\begin{equation}
    \mu(t_L)= \mu_o \left [ 1- \frac{\phi_a(t_L)}{\phi_m}\right ]^{-[\eta]\phi_m}
    \label{Eq:mod_KD-model}
\end{equation}
where $\phi_a$ is the apparent volume fraction due to aggregation, given by:
\begin{equation}
    \phi_a=\phi \left[\frac{r_a}{r_p}\right]^{3-D_f}
\end{equation}

The modified Krieger–Dougherty model, presented in \autoref{Eq:mod_KD-model}, incorporates the effects of particle agglomeration within the bulk liquid to estimate the time-dependent viscosity of the droplet, $\mu(t_L)$. 

However, the value of $k_{agg,pr}$ obtained using \autoref{eq:peri} significantly overestimates the aggregate size ($r_a$), yielding unrealistic values over the timescale relevant to the present experiments. This indicates that even though diffusion plays a dominant role in agglomeration ($Pe_{ortho}$), the perikinetic agglomeration model does not include effects like re-dispersion or breakup of clusters, homogeneous aggregation, temporal increase of crowding, steric hindrance or other repulsive forces, which might cause the over-estimation of aggregation.  

Thus, the perikinetic aggregation model is modified  \citep{oyegbile2016flocculation} to incorporate the parameter $\alpha$ (collision efficiency factor or sticking probability), which accounts for the fact that not all collisions result in aggregation, thereby relaxing the idealised assumption used in by Smoluchowski's perikinetic aggregation model:
\begin{equation}
    k_{agg,pr,mod}=\alpha\frac{8k_BT}{3\mu}
    \label{eq:peri_mod}
\end{equation}
where the value of $\alpha \ll 1$ for bulk liquid in the TM-10 nanofluid droplet in current experiments \citep{mohtaschemi2014rheology,gregory2009monitoring}. 

\begin{figure}
    \centering
    \includegraphics[width=0.8\textwidth]{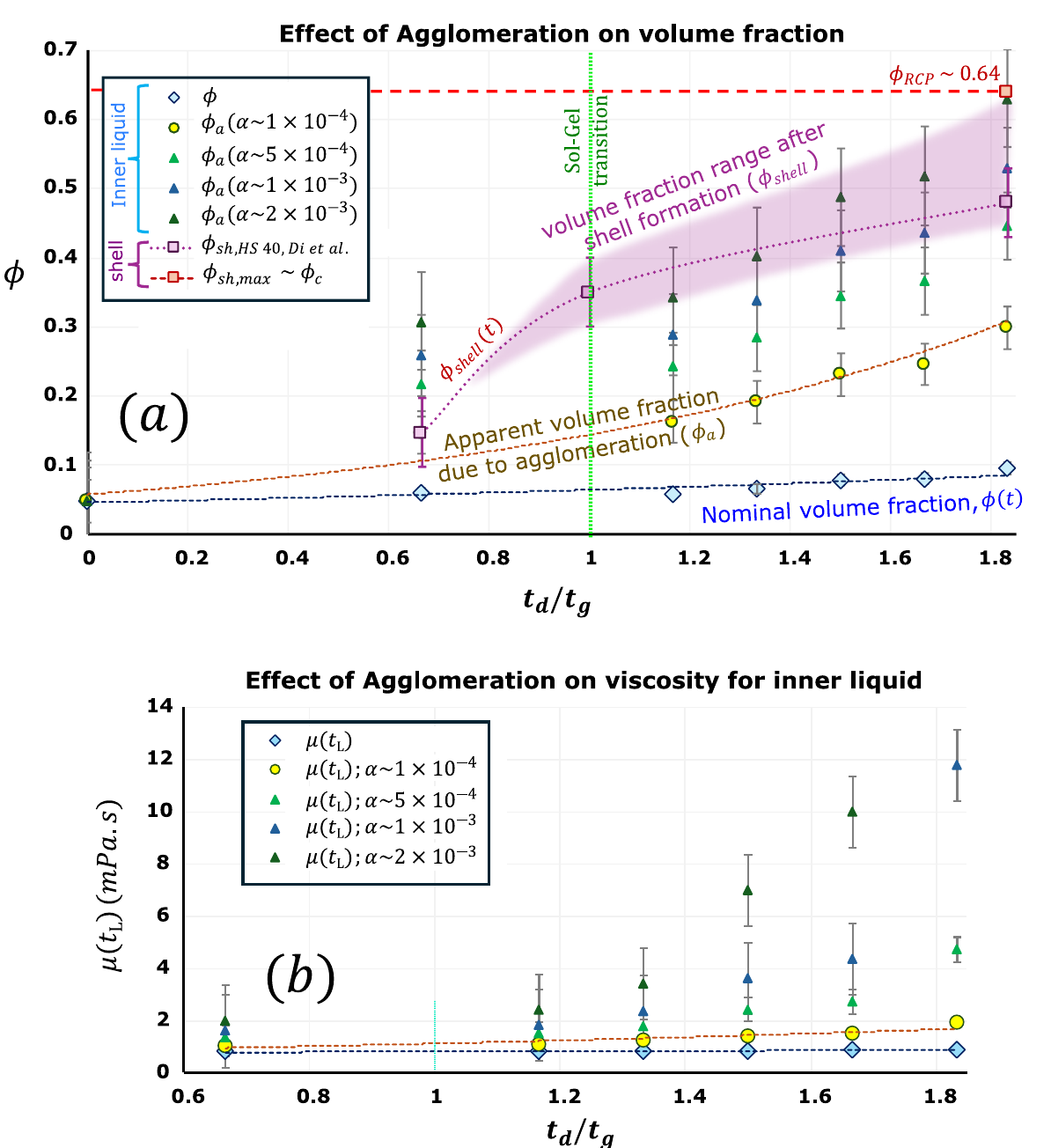}
    \caption{(a) Estimate of the variation of volume fraction during the evaporation of TM-10 (for bulk liquid and shell): $\phi$, $\phi_a$ at different values of $\alpha$ (collision efficiency factor), $\phi_{sh,HS-40}$\citep{di2012rheological} and $\phi_{RCP}\sim 0.64$. (b) Viscosity variation, $\mu(t_L)$ estimated using \autoref{Eq:KD-model} and \autoref{Eq:mod_KD-model} for the bulk medium.}
    \label{fig9p5:Phi_Visc}
\end{figure}

Thus, to obtain the estimate for the viscosity variation, the temporal evolution of $\phi_a$ in the interior of the evaporating TM-10 droplet was evaluated for various $\alpha \ll 1$ and compared with the nominal volume fraction $\phi(t_L)$, as shown in \autoref{fig9p5:Phi_Visc}a. The blue dotted line and blue rhombus markers indicate the temporal variation of the volume fraction ($\phi$) of the inner liquid during TM-10 droplet evaporation when agglomeration is not considered. On the other hand, the yellow dotted line with circular markers depicts the temporal evolution of the apparent volume fraction of the inner liquid ($\phi_a$) during evaporation for the case $\alpha \sim 1\times10^{-4}$. This shows a marked deviation from $\phi(t_L)$, with $\phi_{a,\max}$ reaching only about $0.3$ over the droplet lifetime, corresponding to the Sol regime \citep{di2012rheological}. In contrast, higher $\alpha \ge 5\times10^{-4}$ values lead to unrealistically high $\phi_a$ in the droplet interior (shown in \autoref{fig9p5:Phi_Visc}a with triangular markers), reaching $0.4$–$0.5$, suggesting gelation. However, this does not agree with experimental observations, which clearly showed liquid-like behaviour of the internal liquid in the scope of the present experiments. All these cases, $\phi$, $\phi_{a,\alpha\sim E-4}$, and $\phi_{a,\alpha \ge 5E-4}$ are correspond to the inner liquid region and are labelled as “inner liquid” in the legend of \autoref{fig9p5:Phi_Visc}a. 

Conversely, estimating the outer-shell volume fraction ($\phi_{shell}$) is more complex, as it depends not only on the agglomeration kinetics described earlier in \autoref{sec:composition}, but also on advective transport from internal circulation and on surface recession, which continuously gathers and concentrates particles at the liquid–vapour interface \citep{tsapis2005onset, saiseau2025skin}. The interplay between these mechanisms governs the rate and extent of particle accumulation at the shell, ultimately determining its local volume fraction and mechanical properties. As discussed earlier, gelation in the shell occurs when the local volume fraction reaches $\phi_g$, while solidification takes place when it approaches the maximum packing fraction $\phi_m$ of the constituent particles \citep{chen2007rheological}. Accordingly, around $t_d \sim t_g$, when the shell begins to form, $\phi_{\text{outer}} \rightarrow \phi_g$, and subsequently, as $\phi_{\text{outer}} \rightarrow \phi_m$, solidification of the outer shell ensues around $t_d \rightarrow t_{sh}$.

\cite{di2012rheological} showed that for colloidal silica (HS-40, $d_{p,HS-40}\sim12\text{nm}$), the gelation occurs around $\phi \sim 0.35$ and solidification of the shell occurs around $\phi\sim 0.51$. However, the same cannot be directly applicable in case of colloidal silica (TM-40 or TM-10) that is used in current experiments, whose particle size is $d_{p,TM-40}\sim d_{p,TM-10}\sim22\text{nm}$. Furthermore, \cite{lu2008gelation} demonstrated that gelation in colloidal particles is governed by the range and strength of interparticle attractions relative to particle diameter. For a fixed absolute attraction range, the relative attraction range (i.e., attraction range normalised by particle diameter) is larger for smaller particles, thereby promoting gelation more readily in nanofluids containing smaller particles. This suggests a relatively higher gelation tendency in HS-40 (characterised by \cite{di2012rheological}), with gelation occurring at lower volume fractions (i.e., a lower $\phi_g$) compared to TM-40, from which the current sample TM-10 is prepared.

This suggests that the critical gelation volume fraction, $\phi_{g,HS-40} \sim 0.35$ (for HS-40), will be an underestimation of $\phi_g$ in the context of the current sample TM-10. Thus, an error bar has been added to $\phi_{shell,HS-40}$ in \autoref{fig9p5:Phi_Visc}a, at $t_d/t_g\sim1$ (sol-gel transition). This also indicates that, similar to that in gelation, the solidification also occurs at a higher $\phi$ in case of TM-10 when compared to the value provided by \cite{di2012rheological} ($\phi_{c,HS-40}\sim0.51$). Moreover, the maximum packing fraction for spherical monodispersed particles (random close packing), \(\phi_m\), is approximately 0.64, which implies that the upper limit of \(\phi_c\) for TM-10 approaches 0.64. Therefore, it can be deduced that, for a fully solidified shell in TM-10, the particle volume fraction lies within the range $0.51 < \phi_c \leq 0.64$. The same has been depicted in \autoref{fig9p5:Phi_Visc}a, with red and purple square markers, and the entire range of $\phi_{shell}$ has been depicted as a pink patch. The two estimates for the volume fraction of the shell: $\phi_{sh,HS-40}$ and $\phi_{shell,RCP}$ are labelled as "shell" in the legend of \autoref{fig9p5:Phi_Visc}a. 

\autoref{fig9p5:Phi_Visc}b presents the results of the approximate scaling for the temporal variation of viscosity, obtained using the different volume fractions shown in \autoref{fig9p5:Phi_Visc}a. The results show that, for the bulk inner liquid, when agglomeration effects are considered (yellow circles), only a slight increase in viscosity is observed. Despite this effect, the order of magnitude remains unchanged, consistent with the experimentally observed liquid-like behaviour of the inner liquid in the present study. However, the viscosity of the shell cannot be estimated theoretically without the volume fraction, packing efficiency, local aggregate size and morphology inside the shell, which is outside the scope of current experiments as it needs further investigation using SEM and rheological characterisation. 

Furthermore, using the literature data in conjunction with the current analysis allows for determining the order of magnitude of the shell’s elastic modulus ($G$), a key parameter governing shell rupture and atomisation in the present study. \cite{cao2011aggregation} reported that, for colloidal silica during gelation ($r_{p,\text{cao et al.}}\sim 7\text{nm}$), the elastic modulus scales as $G \sim \phi^{3.3}$, consistent with the theoretical scaling $G \sim \phi^{\frac{3}{3-D_f}}$ (\citep{shih1990scaling,wu2001model}). Furthermore, the micromechanical network model by \cite{zaccone2011approximate} showed that the shear modulus is directly proportional to the number density, which implies that at a given $\phi$, $G \sim r_p^{-3}$. Thus, using order of magnitude of $G$ for HS-40 available in the literature \citep{di2012rheological}, an approximate order of magnitude of $G$ can be obtained using $\frac{G_{HS}}{G_{TM}} \sim \left( \frac{r_{p,HS}}{r_{p,TM}} \right)^{-3}
$. Hence, the estimated order of magnitude of the elastic modulus for the gelatinous shell in TM-10 is $\sim \mathcal{O} (10^3-10^4)Pa$.

\FloatBarrier
 \subsection{Atomisation dynamics at different shock delays ($t_d$)}\label{sec:atomz}

\begin{figure}
    \centering
    \includegraphics[width=1\textwidth]{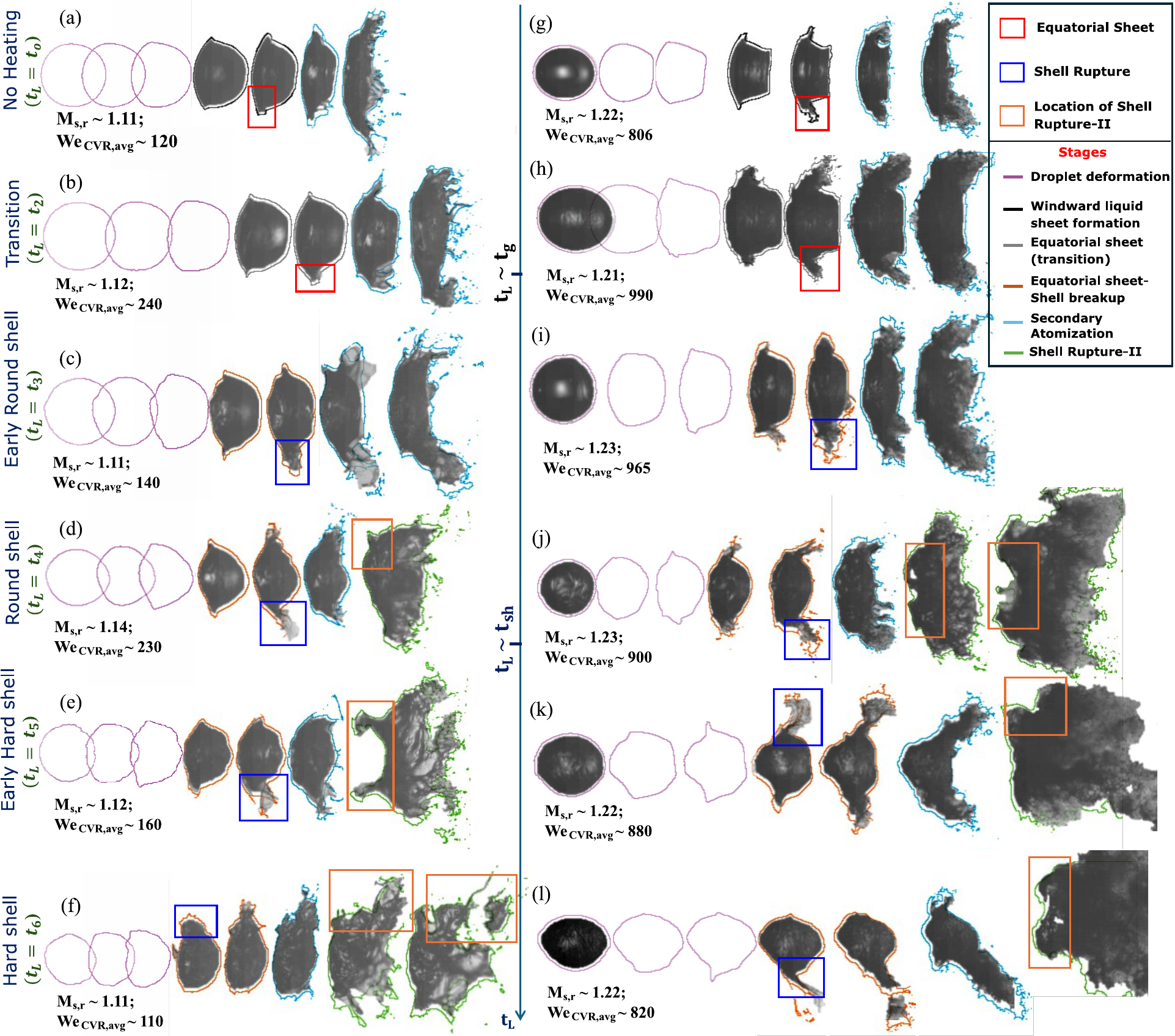}
    \caption{Topological evolution of TM-10 droplet during aero-breakup correponding to the shock strengths: (a-f) $M_{s,r}\sim 1.12$ (5kV) and (g-l) $M_{s,r}\sim1.22$ (8kV). The time-series evolution images at the same vertical position on either side of the centreline i.e., (a,g), (b,h), etc. correspond to same $t_d$. The shock flow direction is from left to right.}
    \label{fig10:topol_TM10_delays}
\end{figure}

The \autoref{fig10:topol_TM10_delays} shows the topological evolution of the droplet shape characteristics during the interaction with the shock flow for different $t_d$, and shock strengths. As explained in \autoref{subsec:global Obs} and \autoref{sec:composition}, with increase in the shock delay ($t_d$), the viscosity increases due to increase in particle concentration and agglomeration for $t_0 \le t_d\le t_2$. Subsequently, for $t_d>t_g$ (sol-gel transition), gelatinous shell forms for $t_d\sim t_3, t_4$ (\textbf{Movie 5 $\&$ Movie 6}) and the shell eventually solidifies beyond $t_d>t_{sh}$ corresponding to shock delays of $t_d\sim t_5, t_6$ (\textbf{Movie 7 $\&$ Movie 8}). The outlines of the droplet shape evolution are colour-coded to indicate the different stages of aero-breakup occurring for different $t_d$ in \autoref{fig10:topol_TM10_delays}. The corresponding droplet images have been overlaid onto the outlines to provide better visual representation. The initial droplet images for each $t_d$ are shown in \autoref{fig10:topol_TM10_delays}g–l. A smooth droplet surface is observed for $t_d < t_2$, corresponding to the steady evaporation regime. In the transition regime $t_d=t_2\sim t_g$; see \autoref{fig10:topol_TM10_delays}h), slight surface texturing begins to appear. Further increase in $t_d>t_2$ leads to the gelatinous shell regime (labelled as “Round shell” in \autoref{fig10:topol_TM10_delays}i,j), where the droplet exhibits pronounced surface texturing along with mild distortion in the droplet’s line-of-sight profile. This effect becomes more pronounced in the solid shell regime (referred to as “Hard shell” in \autoref{fig10:topol_TM10_delays}k,l), where the droplet undergoes significant shape deformation and severe surface texturing, indicating the formation and subsequent deformation of a solid shell.

It can be observed that during the initial interaction with $\rm v_s$ (green region in \autoref{fig8:timescales_Stages}b), the initial deformation is observed in all cases, and is highlighted using a pink outline in \autoref{fig10:topol_TM10_delays}. As discussed in \autoref{sec:Int_timescales}, after the CVR reaches the droplet at $t=t_{CVR}$, the droplet atomisation ensues. Since the $We_{CVR,avg}$ associated with the flow is $>100$, shear-induced entraninment (SIE) mode of atomisation is dominant in all the cases. This results in the formation of a sheet from the liquid on the windward side of the droplet (indicated by red rectangles in \autoref{fig10:topol_TM10_delays}a,b,g,h) for lower shock delay cases ($t_d\le t_2$). These stage of initial sheet formation and extension are represented using black outline in \autoref{fig10:topol_TM10_delays}a,g. The sheet formed subsequently undergoes secondary atomisation through the formation of ligaments (shown using blue outline in \autoref{fig10:topol_TM10_delays}). At higher shock strengths ($M_{s,r}$), the KH waves are observed to cascade and aggravate on the windward side of the sheet.

\begin{figure}
    \centering
     \includegraphics[width=1\textwidth]{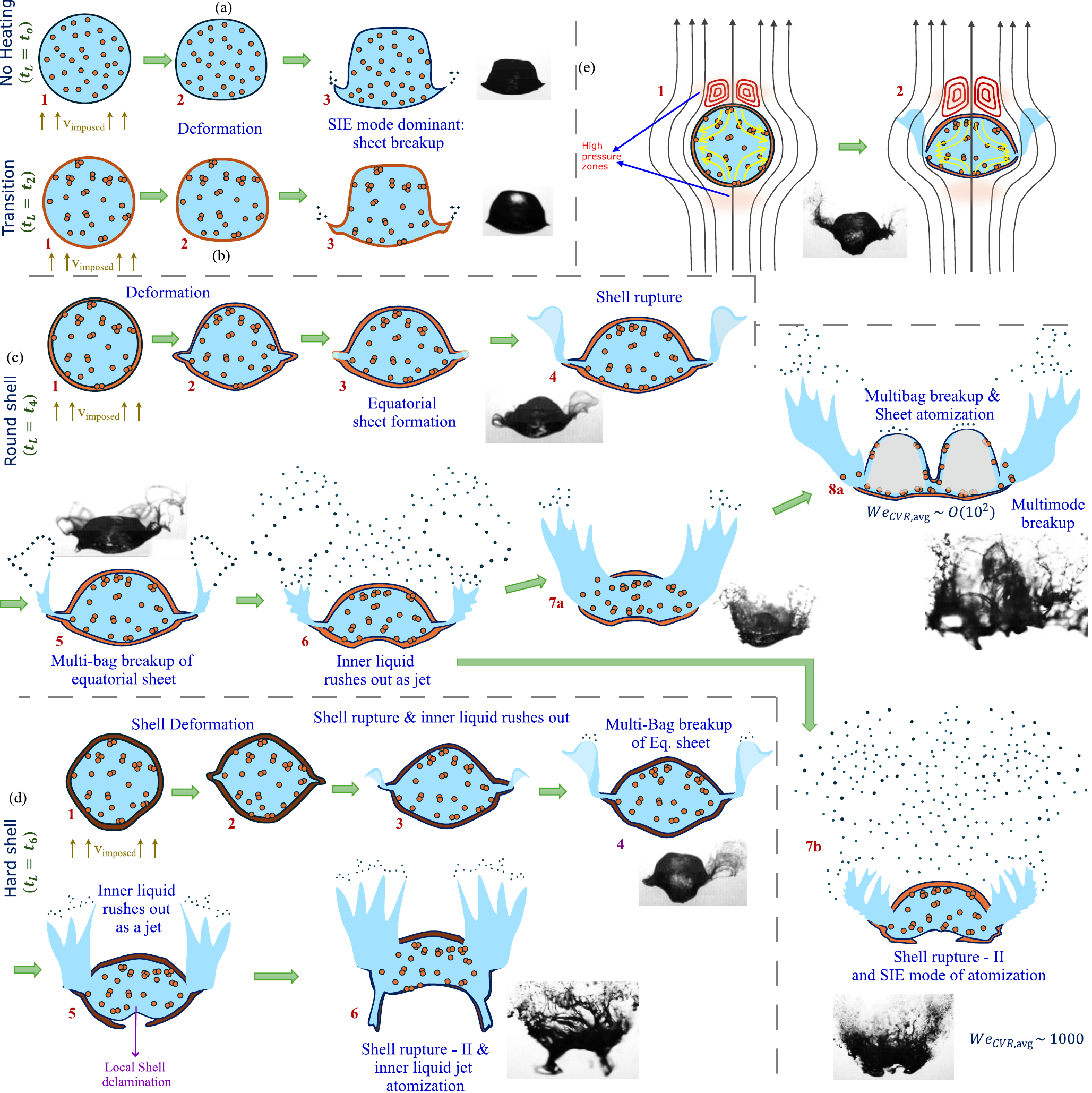}
    \caption{Schematic illustrating atomisation regimes with emphasis on shell rupture dynamics, showing the sequence of events and an overview of the different shell rupture and atomisation modes during interaction for various regimes: (a) $t_d = t_0$ (no heating), (b) $t_d = t_2$ (transition), (c) $t_d = t_4$ (gel shell), and (d) $t_d = t_6$ (solid shell).}
    \label{fig11:Atomz_Schematic}
\end{figure}

Interestingly, near the transition regime ($t_d\sim t_2\sim t_g$), even though the droplet behaviour is still in the sol-phase (liquid-like similar to $t_d<t_2$), the windward face of the droplet tend to show slight inflexion in its curvature during sheet formation (grey outline in \autoref{fig10:topol_TM10_delays}b,h) unlike monotonic curvature for $t_d<t_2$. The same has been depicted in the schematic in \autoref{fig11:Atomz_Schematic}b,c(2-3), showing non-monotonic curvature of the windward surface near and after sol-gel transition ($t_d\ge t_g$). This effect intensifies with further increase in shock delays ($t_3\le t_d< t_5$) beyond the sol-gel transition ($t_L>t_g$). After significant gelatinous shell is formed at the droplet surface ($t_d>t_g$), the sheet formation is significantly altered. As depicted by the orange outline in \autoref{fig10:topol_TM10_delays}c,d,i,j, the sheet tends to form more from the central/equatorial plane than from the windward face of the droplet. This can be explained using the liquid transport inside the droplet due to the formation of high-pressure zones near the forward and rear stagnation during shock-droplet interaction, as established by \cite{sharma_shock_2021,sharma_shock-induced_2023}. 

This induces a net transport of liquid from the poles towards the equatorial region, as illustrated in \autoref{fig11:Atomz_Schematic}e, leading to fluid accumulation near the equatorial plane. The accumulated liquid exerts pressure on the gelatinous shell, attempting to pierce through it as protrusions (indicated by orange outline in \autoref{fig10:topol_TM10_delays}c,d,i,j) and escape outward as shown in \autoref{fig11:Atomz_Schematic}c(2-4). This leads to the formation of an equatorial sheet (protrusion) near the central plane, and while they extrude out, Rayleigh–Taylor (RT) instabilities get modulated over them. This result in the bag-on-sheet mode of multimode breakup, which occurs due to the formation of multiple bags within the sheet due to RT piercing, as illustrated in \autoref{fig11:Atomz_Schematic}c(4). Once the bags breakup, the viscous layer (gel-shell) ruptures (as indicated by blue rectangles in \autoref{fig10:topol_TM10_delays}c,d,i,j), the inner liquid starts to rush outside in the form of a jet/sheet which undergo secondary atomisation through ligament formation. This is evident from \autoref{fig7:GlobalObv_TM10Delay}b at 1.60ms. This is also depicted in \autoref{fig11:Atomz_Schematic}c(5,6,7a), where after the shell gelatinous punctures, the inner liquid rushes out, undergoing atomisation, which is followed by further disintegration of the gelatinous shell around the droplet. Experimental observations also reveal that sheet or protrusion formation occurs repeatedly in multiple cycles at different locations on the windward face. These events take place wherever the internal liquid is forced through a weakened or perforated region of the gelatinous shell, resulting in localised extrusion or sheet ejection. This behaviour is evident from the jet- or sheet-like protrusion piercing the outer shell, as shown in \autoref{fig7:GlobalObv_TM10Delay}b  at 1.48ms.

Simultaneously, the shell near the forward stagnation point undergoes delamination-like phenomenon due to due to the local external shear as indicated by orange rectangle in \autoref{fig10:topol_TM10_delays}d,j which result in splitting of the windward droplet surface and thereby the fracture of the forward surface of the gelatinous shell into two branches, thus exposing the inner liquid, leading to further atomisation. This event is hereby termed as "shell rupture-II" and is shown using a green outline in \autoref{fig10:topol_TM10_delays}, which occurs due to the local external shear near the forward stagnation point and liquid movement towards the equatorial sheet. When the underlying liquid is gradually transported away from the location, the thin shell which is weakly bonded to the underlying liquid detaches, resulting in delamination - shell rupture-II. After the shell rupture-II event in this regime, at high shock strengths ($We_{CVR,avg}\sim 1000$), SIE atomisation mode ensues, resulting in total disintegration of the droplet. However, at low shock strength ($We_{CVR,avg}\sim 100-200$) post-shell rupture-II event as the central liquid expands, Rayleigh-Taylor (RT) piercing occurs at the centre, as shown in \autoref{fig7:GlobalObv_TM10Delay}a,b ($t_d=t_3$ at 2.15ms, $\& \ t_4$ at 1.83ms). 

A similar behaviour has been observed for high shock delay beyond the solidification limit ($t_d\ge t_s$), where a thin layer of the gelatinous shell solidifies due to overcrowding of the particles and aggregates at the droplet surface, as discussed in \autoref{sec:composition}. Because of the formation of the solid shell, the droplet surface is observed to be irregular, rough and deformed. Due to the stochasticity in the process, the shell formation phenomenon is not uniform at all locations. This results in a deformed/partially buckled solid shell due to acoustic forces, for high shock delay cases ($t_d=t_6$). During the aerobreakup of the solid shell regime droplet, the shell exhibits significant resistance to the droplet deformation and atomisation, resulting in minimal shell deformation, shown in \autoref{fig10:topol_TM10_delays}e,f,k,l, followed by shell puncture at equatorial locations. The similar to gelatinous shell regime, a shell protrusion extrudes out near the equatorial location, which gets punctured eventually to release the inner liquid (blue rectangle in \autoref{fig10:topol_TM10_delays}e,f,k,l). Since, the solid-shell has higher elastic modulus and strength compared to the gelatinous shell, the shell puncture at the equatorial location only occurs when the internal pressure buildup overcomes the hoop stress offered by the solid shell. Once, the puncture occurs in the solid shell, the inner liquid rushes out, followed by peeling-off and fragmentation of the solid brittle shell from the droplet surface, as seen in \autoref{fig7:GlobalObv_TM10Delay}d $>1.95ms$. The solid shell formation significantly reduces the droplet atomisation tendency which also results in the presence of large lumps of shell fragmenting and flying-off during the aerobreakup.

It is interesting to note that a consistent trend is followed by the droplet shape variation during the initial deformation phase when subjected to the shock flow, as indicated by the pink outlines in \autoref{fig10:topol_TM10_delays}a–f/g-l. For the unheated and steady stage regimes, during the initial deformation, the droplet showed a cup-cake shape whose windward face characterised by a monotonic and smooth curvature extends outward forming a sheet, similar to previous studies on non-evaporating droplets by \cite{sharma_shock_2021,chandra_shock-induced_2023,sharma_shock-induced_2023}. However, as the shell begins to form, the windward face begins to show inflexion in its curvature, showing sharper features, which result in the formation of an equatorial sheet from the central plane (orange outlines in \autoref{fig10:topol_TM10_delays}). The shell punctures locally at the equatorial locations, resulting in the release of inner liquid. As the shell solidifies, the deformation is also suppressed, offering higher resistance to the local shell puncture.

\FloatBarrier
\subsubsection{Droplet breakup modes} 
In \autoref{sec:atomz}, the effect of shell formation on the atomisation mechanism have been investigated. The results showed that in the current experiments, the droplet atomisation follows different breakup modes, depending on the conditions such as gelation, shell formation and shock strength. Apart from Rayleigh-Taylor piercing (RTP) and shear-induced entrainment (SIE) modes of atomisation, the nanofluid droplet showed other modes of atomisation. Due to shell formation, the deformation and atomisation of the droplet are resisted by the outer shell, leading to a slower droplet response, as evident from \autoref{fig7p5:DropletDef_TM10_Del}. During the interaction, the viscous outer shell (for $t_g<t_L<t_{sh}$) initially deforms and shows an equatorial protrusion as the inner liquid is forced through the outer shell. This occurs as the internal liquid (still fluid) is forced through a weakened or perforated gelatinous shell, which forms a localised extrusion or sheet ejection, which may resemble a jet, bag, or finger that pierces the shell. This leads to the formation of an equatorial sheet, upon which Rayleigh–Taylor piercing occurs, resulting in a bag-on-sheet mode of atomisation. Once the outer layer of viscous gel-shell ruptures, the inner liquid rushes outside as a jet, undergoing secondary atomisation. 

\begin{figure}
    \centering
     \includegraphics[width=1\textwidth]{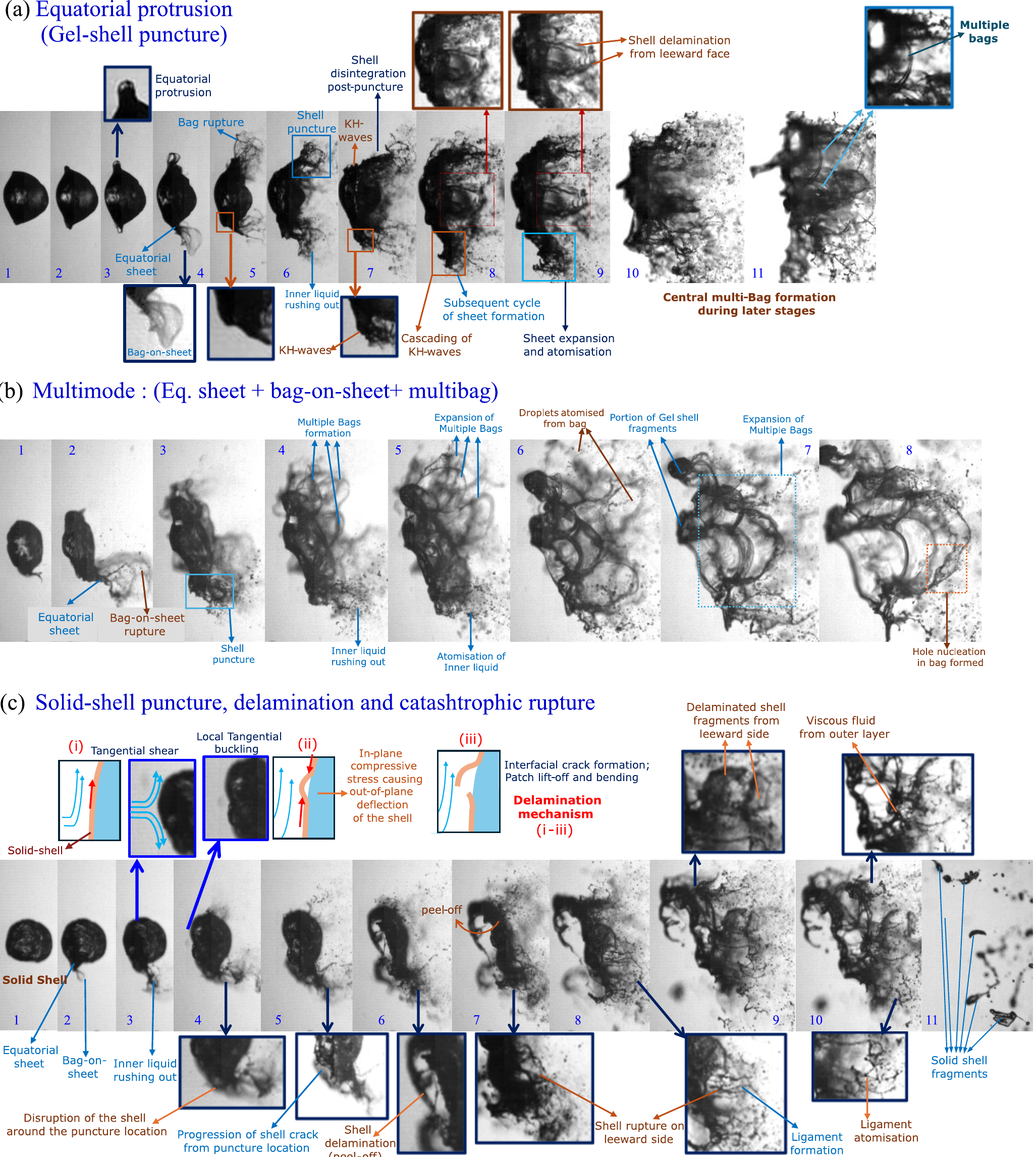}
    \caption{Mechanism of different modes of breakup are illustrated: (a) Breakup via Equatorial protrusion, (b) Multimode breakup: Eq. sheet + bag-on-sheet+multibag, (c) Solid-shell puncture, delamination and catastrophic rupture.}
    \label{fig12:Atomz_Modes1}
\end{figure}

\autoref{fig12:Atomz_Modes1}a depicts the mechanism of the Equatorial protrusion followed by Gel-shell puncture during atomisation. Following shell puncture, the shell begins to disintegrate due to the loss of structural integrity, allowing the inner liquid to escape, as shown in \autoref{fig12:Atomz_Modes1}a(5–7). The escaping jet of inner liquid undergoes atomisation in a manner analogous to that of a liquid-jet in a crossflow. Simultaneously, Kelvin-Helmholtz instability is observed on the windward face of the droplet as shown in \autoref{fig12:Atomz_Modes1}a(7). The KH–waves evolve progressively, cascading over time and transporting liquid outward on the windward face of the droplet. This results in the formation of a liquid sheet, which subsequently grows and expands due to continued aerodynamic forcing. Eventually, the thinning sheet becomes unstable and undergoes atomisation, breaking up into ligaments and droplets. As shown in \autoref{fig12:Atomz_Modes1}a(8), multiple cycles of KH-wave growth and sheet formation occur during the droplet atomisation. The formation of KH waves is similar to that observed in the atomisation of pure liquid droplets. However, in the case of evaporating nanofluid droplets, the presence of a shell gives rise to a significantly different atomisation mechanism, which is characterised by equatorial protrusions and shell puncture. The shell puncture can be analysed by evaluating the hoop stress ($\sigma_{\theta}$) resistance of the shell against the pressure exerted by the inner liquid as it tries to escape outside. The hoop stress for the droplet shell can be evaluated using $\sigma_{\theta}\sim \frac{\Delta P_{in} D}{4 \delta_{sh}}$, where $\Delta P_{in}$, $D$ and $\delta_{sh}$ are pressure difference across the shell pressure, droplet diameter and shell thickness, respectively. For puncture-type failure ($\sigma_{\theta} \sim G$), the critical internal pressure difference required to cause rupture can be estimated using $\Delta P_{in}\sim 4\delta_{sh}G/D$. Here, only the approximate range of the elastic modulus \((G)\) is evaluated in \autoref{sec:composition}; however, the detailed variation of \(G\) and the shell thickness \((\delta_{\mathrm{sh}})\) remains unknown and lies beyond the scope of the present experiments, requiring more in-depth investigation through rheological measurements and SEM imaging.

\FloatBarrier

\autoref{fig12:Atomz_Modes1}b shows the multimode breakup observed post-shell rupture. Initially, the bag-on-sheet mode of breakup occurs on the equatorial sheet, which results in shell rupture. Subsequently, at low shock strengths ($M_{s,r}\sim1.1$), Rayleigh–Taylor piercing (RTP) is observed during the later stages of interaction with $\rm v_{CVR}$, leading to the formation of multiple bags, as shown in \autoref{fig12:Atomz_Modes1}b. Finally, hole nucleation occurs within the expanding bags as they stretch and become progressively thinner, resulting in secondary atomisation. \autoref{fig12:Atomz_Modes1}b(6-8) also illustrates that portions of the outer gelatinous shell form stamen-like structures during the multiple bag mode of atomisation, while the majority of secondary atomisation arises primarily from hole nucleation. \autoref{fig12:Atomz_Modes1}a(11) further illustrates the multiple bag mode of atomisation caused by Rayleigh–Taylor (RT) piercing, as the droplet continues to deform and flatten after shell rupture. This multiple bag formation due to Rayleigh–Taylor piercing is observed exclusively in the gelatinous shell regime ($t_L\sim t_g$), where a substantial volume of the inner liquid remains and the outer shell has not yet solidified. 

\begin{figure}
    \centering
     \includegraphics[width=1\textwidth]{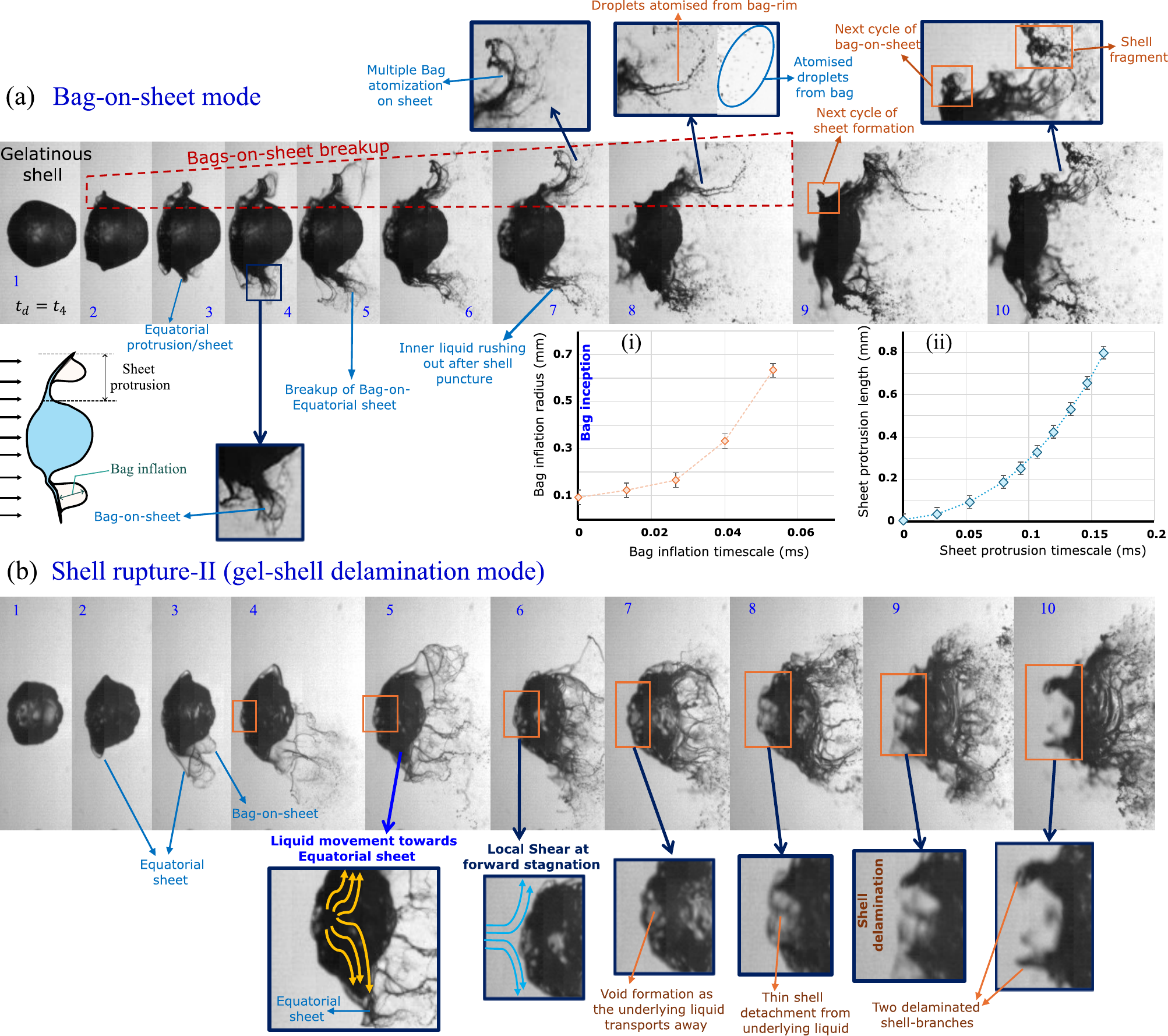}
    \caption{Mechanism of different modes of breakup are illustrated: (a) Bag-on-sheet mode of breakup, (b) Shell-rupture-II (delamination mode of breakup).}
    \label{fig13:Atomz_Modes2}
\end{figure}

Although solid shell puncture, bag-on-sheet breakup, and liquid jet escape also occur in the solid shell regime similar to the gelatinous shell regime, additional shell breakup modes emerge once the shell begins to solidify, i.e.,  $t_g < t_L \le t_{sh}$, as shown in \autoref{fig12:Atomz_Modes1}c and \autoref{fig13:Atomz_Modes2}a,b. The breakup modes such as catastrophic solid shell puncture and shell rupture-II (see \autoref{fig10:topol_TM10_delays}) are observed in solid shell regime, as discussed in \autoref{sec:atomz}. \autoref{fig12:Atomz_Modes1}c illustrates the puncture of the solid shell following initial deformation, equatorial sheet formation, and bag-on-sheet breakup, all of which are comparable to those observed in the gelatinous shell regime (\autoref{fig12:Atomz_Modes1}a). Since the outer shell has solidified for $t_L\sim t_{sh}$, it exhibits a more brittle nature during puncture, with relatively greater resistance to deformation and equatorial sheet formation. This behaviour is evident in \autoref{fig7:GlobalObv_TM10Delay}d,g,h, where equatorial protrusions occur stochastically on only one side or both sides of the droplet after solidification, owing to the increased resistance to shell rupture. A similar observation is made in \autoref{fig12:Atomz_Modes1}c, where shell puncture and equatorial sheet formation are restricted to one side, attributed to the presence of the solid shell.

In addition, for $t_L\sim t_{sh} \sim t_6$, shell puncture leads to the initiation of a crack in the solid shell, which subsequently propagates and disrupts the solid shell structure, leading to delamination or peel-off, as shown in \autoref{fig12:Atomz_Modes1}c(6–8). Delamination of a thin solid shell on a liquid droplet is an interfacial fracture process initiated by local pressure differences or tangential shear loading, which lift a finite patch of the shell away from the liquid core \citep{hutchinson1991mixed}. In the buckle-driven delamination mode described by \cite{hutchinson1991mixed}, in-plane compressive hoop stresses, which may arise from aerodynamic tangential shear, induce local buckling of the shell in the tangential direction. This buckling generates an out-of-plane deflection that concentrates stress at the interface, nucleating an interfacial crack. Once the crack forms, aerodynamic forces acting tangentially to the droplet surface promote its growth, while also bending and lifting the detached patch into the airstream. As the delaminated region extends, progressive loss of adhesion at the blister base (anchor zone) leads to complete peel-off of the shell from the droplet surface. This mechanism has been shown in \autoref{fig12:Atomz_Modes1}c(i,ii,iii) using both zoomed-in experimental images and schematics. It is to be noted that this is different to the mode of buckling observed in the literature in nanofluid shells (under external pressure in normal direction).

Thus, through the above mechanism, the solid shell lift off from the droplet due to external aerodynamic forces after the solid shell separation delamination is shown in \autoref{fig12:Atomz_Modes1}c(6-8). This results in the disruption of the solid shell on the windward side of the shell. Furthermore, shell disruption is also observed on the leeward side of the droplet, leading to delamination and/or catastrophic rupture of the solid shell, as shown in \autoref{fig12:Atomz_Modes1}c(7–8). This rupture is followed by the atomisation of the escaping inner liquid through ligament formation and subsequent secondary breakup. The presence of the solid shell and its rupture is also indicated by the solid rigid fragments from the solid shell visible in \autoref{fig12:Atomz_Modes1}c(11), post-interaction.

On the other hand, for $t_g < t_L < t_{sh}$, which is the early solid-shell regime, the solidification has occurred non-uniformly in the gelatinous shell, wherever the local volume fraction reaches $\phi \rightarrow \phi_s$. Delamination mode of breakup is also observed in this regime, where the high-pressure near the forward stagnation point results in advection of fluid away towards the equatorial sheet. In addition to it, the local shear near the forward stagnation point leads to the disruption of the thin shell layer, resulting in a crack formation \citep{unverfehrt2015deformation,chang1993experimental}. Thus, the combined effect of the underlying liquid transport and the shell cracking results in the formation of a void and shell delamination,as shown in \autoref{fig13:Atomz_Modes2}b(7-10). This has also been indicated using orange rectangles in the \autoref{fig13:Atomz_Modes2}b, where the outer shell, post-rupture, parts into two branches, which delaminate from the liquid surface, and thus, exposing the inner liquid (see \autoref{fig10:topol_TM10_delays}, green outline). This process has been termed as "shell rupture-II", which has been discussed in \autoref{sec:atomz}. This mode of breakup (shell rupture-II) occurs during later stages of interaction in gelatinous shell regime ($t_d \sim t_4,t_5$), which leads to the complete disintegration of the droplet.

The bag-on-sheet is observed in this regime as well, where RTP mode is modulated along the equatorial sheet protruding out, after the initial deformation (see \autoref{fig13:Atomz_Modes2}a). The multiple bags formed over the equatorial sheet expand due to aerodynamic forces, which eventually undergo atomisation due to hole nucleation. The schematic shown in \autoref{fig13:Atomz_Modes2}a(below,left) shows the formation and inflation of bags over the equatorial sheet. The temporal variation of the radius of bags-on-sheet (during inflation) and length of the extruded-equatorial sheet has been plotted in \autoref{fig13:Atomz_Modes2}a(i,ii). They show a monotonic increase of bag radius and sheet length, which is consistent among different cases. The detailed zoomed-in snapshots of bags-on-sheet mode of breakup is shown in \autoref{fig13:Atomz_Modes2}a, depicting repeated cycles of equatorial sheet formation and bags-on-sheet breakup. It is also evident from \autoref{fig13:Atomz_Modes2}a(6-9) that the inner liquid is released outside only after the bags-on-sheet breakup, which causes the local puncture of the gelatinous shell. Thus, the different breakup modes observed during the aerobreakup of an evaporating nanofluid droplet have been uncovered in current experiments.

\FloatBarrier
 \subsection{Regime Map}
 
\begin{figure}
    \centering
    \includegraphics[width=1\textwidth]{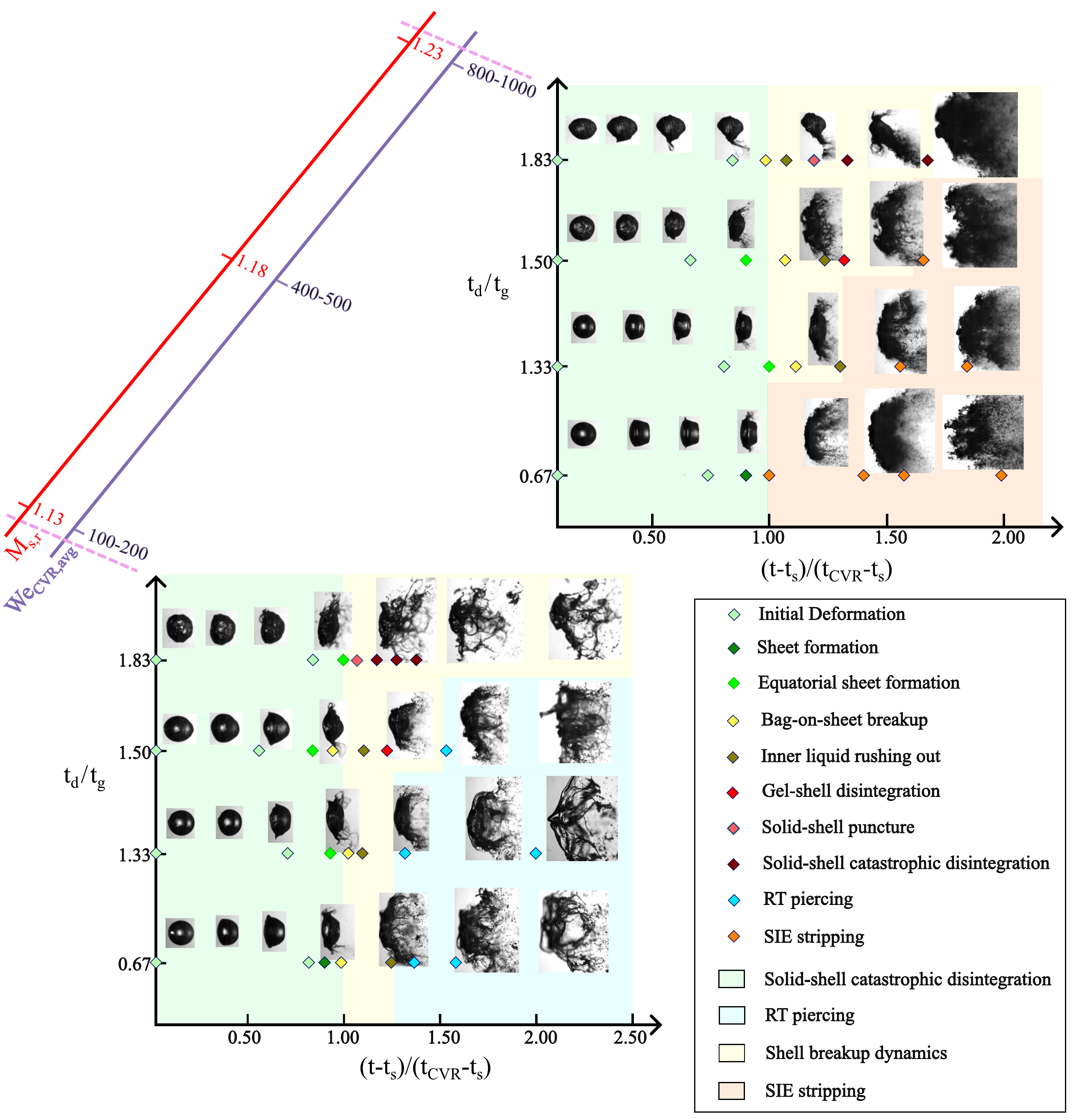}
    \caption{Comprehensive regime map illustrating various atomisation events (normalised atomisation timescale on the x-axis) using coloured markers, alongside broad behavioural regimes during interaction (indicated by background shading). The regime map spans shock delays $t_1 \le t_d \le t_6$, encompassing all regimes during nanofluid evaporation (normalised $t_d$ on the y-axis), and shows the effect of shock strength, given by $M_{s,r}$ or $We_{CVR,avg}$ (z-axis).}
    \label{fig14_Regimemap}
\end{figure}

The various stages and regimes of atomisation, corresponding to different shock delays ($t_d$) and shock strengths ($M_{s,r}$) observed in the present experiments, are compiled and summarised in the regime map shown in \autoref{fig14_Regimemap}. The breakup timescale ($t$) is plotted along the x-axis and is normalised using the shock arrival time ($t\sim t_s$) and CVR arrival time ($t\sim t_{CVR}$), as given by: $t_{norm} \sim \frac{t-t_s}{t_{CVR}-t_s}$. This scaling captures the multistage flow-droplet interaction dynamics, as $\rm v_s$ is only imposed at the droplet for $t>t_s$ and $\rm v_{CVR}$ is imposed on the droplet after $t>t_{CVR}$. This means, for $0<t_{norm}<1$, only $\rm v_s$ interacts with the droplet, while CVR interaction occurs for $t_{norm}>1$. On the other hand, the shock delay ($t_d$), which corresponds to different regimes of shell formation, is plotted along the y-axis and is normalised by the gelation timescale ($t_L\sim t_g$). Thus, $t_d/t_g<1$ corresponds to steady evaporation regime and $t_d/t_g>1$ corresponds to shell formation regimes, i.e., $t_d/t_g\sim 1.33,1.50,1.83$ correspond to $t_d\sim t_3$ (transition), $t_d\sim t_4$ (gelatinous shell) and $t_d\sim t_6$ (solid shell) regimes, respectively. The shock strength is plotted along the z-axis as $M_{s,r}$ and Weber number corresponding to CVR ($We_{CVR,avg}$). Furthermore, representative droplet images corresponding to various events during the interaction are shown in \autoref{fig14_Regimemap}, marked using colour-coded data points (see legend).

The different regimes of breakup behaviour are shown using different coloured backgrounds in \autoref{fig14_Regimemap}. For all the cases, the initial deformation phase where the droplet interaction occurs with $\rm v_s$ is represented by green background ($\frac{t-t_s}{t_{CVR}-t_s}<1$). Following the initial deformation i.e., $t_{norm}>1$, various modes of droplet shell rupture are observed to occur, which are collectively referred to as shell breakup dynamics (indicated by the yellow background). Beyond the yellow region, two distinct types of behaviour are broadly observed. At lower shock strengths ($We_{CVR,avg}< 200$), Rayliegh-Taylor piercing (RTP) induced multiple bag breakup is observed (represented by blue background), whereas at high shock strengths shear-induced stripping is observed (indicated by orange background).

Initially, during $\rm v_s$ interaction phase (green region), the droplet undergoes deformation (light-green markers), and at higher Mach numbers, Kelvin–Helmholtz (KH) waves are also observed. These KH instabilities induce liquid transport, leading to the formation of a sheet (dark-green markers). Moreover, after gelation ($t_d/t_g>1$), the formation of an equatorial sheet protrusion through the shell is observed, represented by the fluorescent green markers. Eventually, for $t_{norm}>1$, the equatorial sheet gets modulated by RT piercing, resulting in bags-on-sheet mode of breakup, indicated by yellow markers. These bags expand and the membrane ruptures through hole nucleation, leading to atomisation. This results in the puncture of the outer shell layer, allowing the entrapped inner liquid to rush out (indicated by olive-coloured markers in \autoref{fig14_Regimemap}). This is succeeded by the secondary atomisation of the escaping liquid, which further disrupts the structural-integrity of the shell. For $t_d/t_g\sim1.5$, the gelatinous shell shows further disintegration and delamination from both leeward and windward sides after the initial puncture (red markers), which is consistent with the literature \citep{grandmaison2021modelling,chang1993experimental}. On the other hand, for $t_d/t_g\sim 1.8$, the solidified shell shows solid-like brittle behaviour, characterised by local shell puncture, delamination of shell in the form of flakes/fragments, and catastrophic disintegration, indicated using dark maroon markers in \autoref{fig14_Regimemap}. All these breakup dynamics involving the gelatinous/solid shell are indicated together as "shell breakup dynamics" using the yellow background in \autoref{fig14_Regimemap}.

At low shock strengths ($We_{CVR,avg}\sim 200$), Gelatinous shell atomisation ($1 < t_d/t_g\le1.5$) progresses through successive stages: beginning with shell puncture, followed by delamination, and culminating in Rayleigh–Taylor piercing-induced multiple bag breakup at the central region of the deformed droplet (blue markers in \autoref{fig14_Regimemap}) after shell rupture. Whereas in the solid-shell regime ($t_d/t_g \sim 1.8$), significant resistance to atomisation is observed due to the presence of a solid shell, which leads to shell deformation, cracking, and crack propagation. This results in catastrophic disintegration of the solid shell, even at low $We_{CVR,avg}$. In contrast, at high shock strengths ($We_{CVR,avg} \sim 1000$), the SIE mode of shear-stripping is observed during the later stages ($t_{norm} > 1$) across all cases.

In the gelatinous shell regime ($1 < t_d/t_g \le 1.5$), following the initial shell rupture and delamination, the shell rupture-II mode of breakup occurs, exposing the inner liquid to the external flow. This is followed by the SIE mode of breakup, resulting in complete atomisation of the droplet. On the other hand, in the solid-shell regime, shell rupture may occur stochastically on one side, either side, or both sides of the droplet, depending on local variations in shell thickness arising from non-uniform solidification. Furthermore, once the solid shell is punctured at an equatorial location, the puncture hole grows progressively, leading to the disintegration of the enclosing shell. The inner liquid jet then undergoes atomisation due to the external flow, followed by the SIE mode of atomisation. The solid-shell rupture mechanisms observed in nanofluid droplets under aerodynamic loading in the present study shows resemblance to the gradual deformation-to-failure pathway shown by \cite{unverfehrt2015deformation,chang1993experimental}.

\FloatBarrier
 \section{Conclusion}\label{sec:conclusion}

This first-of-its-kind study examines how transient shell formation influences shock-induced atomisation of acoustically levitated, evaporating TM-10 nanofluid droplets. Evaporation is initiated by a $CO_2$ laser at $t_L = 0$, and after a delay $t_L = t_d$, a shock tube generates a blast wave via wire explosion at $t = 0$. The blast wave imposes an initial velocity jump, followed by a decaying velocity ($\mathrm{v_s}$), and is later followed by interaction with a compressible vortex ring (CVR) at $t \sim t_{CVR}$ with velocity $\mathrm{v_{CVR}}$. Varying $t_d$ allows the shock to interact with droplets at different evaporation stages, broadly classified as: (i) steady evaporation with $D^2$-law regression and minimal nanoparticle agglomeration; (ii) gel-shell phase, where a porous gelatinous layer forms via sol–gel transition, impeding evaporation ($t_L \ge t_g$); and (iii) solid-shell phase, where the shell reaches maximum packing fraction and solidifies ($t_L \ge t_{sh}$). Each regime exhibits distinct atomisation responses to the shock–flow interaction based on the evolving shell morphology.

Nanoparticle concentration rises during evaporation, promoting agglomeration and bulk viscosity growth, while surface accumulation results in the progressive formation of a gelatinous shell at the droplet surface. In the steady-evaporation regime ($t_d/t_g < 1$), $\mathrm{v_s}$ induces deformation and Kelvin–Helmholtz wave growth, followed by CVR interaction–induced Rayleigh–Taylor (RT) piercing or shear-induced entrainment (SIE) at low and high shock strengths, respectively. In the gel-shell regime ($1 < t_d/t_g \le 1.5$), the viscous–elastic shell resists deformation, producing equatorial sheet protrusion and bag-on-sheet breakup. Shell puncture releases inner liquid as secondary jets, weakening the shell integrity and causing delamination. During the later stages of interaction, at low $We_{CVR,avg}$ ($\sim 100$–$200$), this evolves into RT piercing–driven multibag breakup, whereas at high $We_{CVR,avg}$ ($\sim 1000$) the delamination is followed by SIE stripping. In the solid-shell regime ($t_d/t_g \sim 1.8$), the rigid shell cracks and fractures in a brittle manner, with puncture holes growing from weak spots caused by non-uniform solidification. This leads to flake-like delamination or catastrophic disintegration, after which the remaining liquid is rapidly stripped by external flow.

The breakup modes observed include equatorial protrusion and puncture, bags-on-sheet breakup, solid-shell fracture and delamination, and multimode multibag breakup due to the shell formation. These are consolidated into a holistic regime map parameterised by interaction time ($t_{norm}$), evaporation stage ($t_d/t_g$), and shock strength ($M_{s,r}$, $We_{CVR,avg}$), providing a framework for multistage flow interaction with temporally evolving multicomponent, multiphase droplets.

This investigation addresses an unexplored regime where evaporation-driven interfacial transitions in nanofluids influence atomisation under shock loading, linking droplet composition, shell evolution, and breakup pathways to the timing and intensity of aerodynamic forcing. The results offer a first-of-its-kind framework and insights regarding the engineering of droplet-based systems in which phase change, particulate loading, and aerodynamic stress are all coupled. Further investigations into the detailed mechanics of shell rupture, supported by advanced rheological and SEM characterisation, together with comprehensive studies of droplet size distribution and its evolution during shell formation, will be crucial for deepening understanding of the atomisation of such multicomponent, multiphase, transient droplet systems.

\section*{Declaration of competing interest} \addvspace{10pt}

The authors have no competing interest to disclose.

\section*{Acknowledgments} \addvspace{10pt}

The authors are thankful to SERB - CRG (CRG/2020/000055) for financial support. S.B. acknowledges funding through the INAE (Indian National Academy of Engineering) Chair Professorship. 

\bibliographystyle{jfm}
\bibliography{jfm-instructions}

\end{document}